\documentclass[useAMS,usenatbib]{mn2e}
\usepackage{epsfig,amssymb,amsmath,rotating,lineno,graphicx,breqn}

\title[interpreting large amplitude X-ray variation]{Interpreting the large amplitude X-ray variation of GRS 1915+105 and IGR J17091$-$3624 as modulations of an accretion disc}

\author[Pahari et al.]{Mayukh Pahari$^{1,3}$\thanks{E-mail: mp@tifr.res.in (MP)}, Ranjeev Misra$^{2}$, Arunava Mukherjee$^{2}$, J S Yadav$^{1}$, S K Pandey$^{3}$\\
$^{1}$ Tata Institute of Fundamental Research, Homi Bhabha Road, Mumbai, 400005, India\\
$^{2}$ Inter-University Centre for Astronomy and Astrophysics, Pune, 411007, India\\
$^{3}$ Pt. Ravishankar Shukla University, Raipur, Chhatisgarh, 492010, India }
\begin{document}

\pagerange{\pageref{firstpage}--\pageref{lastpage}} \pubyear{2013}

\maketitle

\label{firstpage}

\begin{abstract}

Using the flux resolved spectroscopy for the first time, we analyse the {\it RXTE}/PCA data of the black hole
X-ray binaries GRS 1915+105 and IGR J17091$-$3624, when both sources show large amplitude, quasi-regular
oscillations in 2.0$-$60.0 keV X-ray light curves ( similar to the $\kappa$ and $\lambda$ classes in GRS
1915+105). For different observations, we extract spectra
during the peak (spectrally soft) and dip (spectrally hard) intervals of the oscillation, and find that
their spectra are phenomenologically complex, requiring at least two distinct
spectral components. Besides a thermal Comptonization component, we find that the disc emission is better modelled by an index-free
multicolour disc blackbody component ({\it p}$-$free disc model) rather than that from a standard accretion disc.
While the peak and dip spectra are complex, remarkably, their difference spectra constructed by treating dip spectra as the background spectra of the peak spectra, can be modelled as a single {\it p}$-$free disc component. 
Moreover, the variability at different time-scales and
energy bands of the peak flux level is always greater than or equal to
the variability of the dip flux level, which strengthens the possibility that the
peak flux level may be due to an independent spectral component added to
the dip one. Using joint spectral analysis of peak and dip spectra with a variable emission component, we verify that the variable component is consistent with {\it p}$-$free disc blackbody and its spectral parameters are similar to that found from the difference spectral analysis. In contrast, we show that for oscillations in the $\theta$ class where soft dips are observed,
the difference spectra cannot be similarly fitted. 
Our result substantiates the standard hypothesis that the
oscillations are due to the limit cycle behaviour of an unstable
radiation pressure dominated inner disc. However, in this
interpretation, the flux variation of the unstable disc can be several
order of magnitudes as expected from some theoretical simulations and need
not be fine tuned to match the factor of $\sim 10$ variation seen
between the peak and dip levels. 
\end{abstract}
  
\begin{keywords} 
accretion, accretion discs --- black hole physics --- X-rays: binaries --- X-rays: individual: IGR J17091$-$3624 --- X-rays: individual: GRS 1915+105
\end{keywords} 

\section{Introduction}

For last 20 years, GRS 1915+105 is the only Galactic micro$-$quasar
showing prolific and complex X-ray intensity morphology. 
Dramatic fall and rise in the
luminosity by an order of magnitude in couple of seconds and 
repetition of such few hundred second fluctuations in cyclic manner over time-scales ranging
from few thousands of seconds to few days, makes GRS 1915+105 unique
with respect to other micro-quasars. 
These drops are most prominent in $\kappa$, $\lambda$ and $\rho$ classes and they are
simultaneous with spectral hardening by a couple of factors
\citep{b5,b16,b99,b55}.

Several attempts have been made to model the spectra of such fluctuations by assuming that the energy spectra at the base/persistent flux
level of the fluctuation are consistent with the low hard state (L$_{bol}$/L$_{EDD}$ $\sim$ 0.05$-$0.2) while energy spectra at the
peak flux level are consistent with the high soft (HS) state (L$_{bol}$/L$_{EDD}$ $\sim$ 0.7$-$1.2) in GRS 1915+105 \citep{b80}. 

The energy spectra at the peak flux level and persistent flux level have been modelled with the disc blackbody (with and without zero-torque conditions) and powerlaw (with and without cut-off)/low temperature Comptonization \citep{b5,b96,b70,b73,b72,b69,b66,b71,b68,b78,b91} or thermal Comptonization of disc blackbody photons along with the presence of a strong non-thermal component/powerlaw tail using {\it RXTE}/PCA, {\it RXTE}/HEXTE \citep{b63,b92,b64} simultaneously with {\it OSSE}/{\it CGRO} observations \citep{b93,b74}, {\it SUZAKU} observations \citep{b95}, {\it BeppoSAX} observations \citep{b57} and {\it INTEGRAL} observations \citep{b66}. \citet{b65} observe that although HS state spectra are dominated by soft thermal emission, it is complex in nature and can not be described by a standard thermal disc. In summary, the spectra of
GRS 1915+105 are complex at different flux levels and there exists a large degeneracy in multi-component spectral fitting.

In a standard {\it steady state} accretion disc, the effective temperature
varies with radii as $T(R) \propto R^{-3/4}$ giving rise to the characteristic
low energy spectral index of $F(E) \propto E^{-1/3}$. A generalization of this
model is when the temperature index is allowed to be a free parameter namely
the p-free disc model where  $T(R) \propto R^{-p}$. In {\tt XSpec}, this model has been
incorporated as {\tt diskpbb} \citep{b67,b90,b85}. There are several possible reasons why
the temperature will deviate from being  $\propto R^{-3/4}$. 
At near Eddington luminosity (L/L $_{EDD} > 0.3 $), the disc is geometrically thicker than the
standard one and there is an
additional cooling mechanism due to advection of energy in the radial direction.
\citep{b89,b90,b11}. The spectra from such slim accretion discs 
can be well approximated as {\it p}$-$free disc spectra with $p \sim 0.5$. Even for lower accretion rate systems,
the spectra may deviate from the standard one. In general, for
optically thick accretion discs, the local effective temperature gets diluted by
a colour factor $\it f$ due to  radiative transfer effects. Although numerical computations suggest that 
the colour factor is nearly independent of radius  with $f \sim 1.7$ \citep{b87}, this
may not always be the case, especially if there is a Comptonizing medium on top
of the disc. If $f$ has a radial dependence, then the observed colour temperature will have a
different radial dependence than the standard case. Another possibility, particularly relevant
to GRS 1915+105, is that for  a non-steady state
disc, the accretion rate may not be constant at different radii, resulting in a deviation of
the temperature profile. Thus the {\it p}$-$free disc model seems to be a reasonable empirical model
to test if it can describe the thermal spectral component of different complex spectra of GRS 1915+105. 

The complex spectra make it difficult to understand the nature
of large amplitude oscillations observed in this source. The disappearance of
a thermal component (thought to be associated with the inner disc)
and the associated jet activities is considered evidence for a
phenomenological picture \citep{b2} which consists of `disc evacuation', jet
creation and subsequent refilling of the disc in a repetitive manner
\citep{b88,b10}. The system is more complex with the soft dips during the $\theta$ class 
which are interpreted as the ejection
of the Comptonized cloud with the disc blackbody and power$-$law being present
\citep{b86}. However, from 
fundamental accretion disc theory, the prime candidate for the cause
of the large amplitude variation  seen in GRS 1915+105 is the radiation pressure 
instability. In the standard accretion disc theory \citep{b62}, the disc is
both secularly and thermally unstable when it is radiation pressure dominated \citep{b89}. Since the inner region of the disc 
is radiation pressure dominated at accretion rates larger than 5\% of the
Eddington value, the standard accretion disc should be unstable and sources
should show large amplitude oscillations. It has been a puzzle why typical
Galactic X-ray binaries, despite showing high luminosities and thermal spectra
from a standard accretion disc, do not show large amplitude variation and seems
not to be subjected to this instability. This could be because the viscosity prescription
is different than that assumed in the standard theory \citep{b21,b50} or 
that stochastic viscous variations stabilize the disc \citep{b90}.   
Notwithstanding that other typical black hole systems do not show
large variations, it is attractive to conclude that the radiation pressure
instability does occur for GRS 1915+105 \citep[e.g.][]{b10,b60,b61}. However, there are several
difficulties or issues regarding this interpretation. GRS 1915+105 shows various
types of variability patterns which seem to be governed by non-linear equations
\citep{b51} and there maybe more than one mechanism responsible for the complex
behaviour. More importantly, some numerical simulations of an unstable radiation pressure
dominated standard disc show different behaviour than what is seen from data \citep[e.g.,][]{b58}. 
\citet{b59} showed that the amplitude of the oscillation in simulations are at least two orders of magnitude
as compared to a factor of 4$-$10 that is observed \citep{b88,b48}. Secondly, their simulations show spike like
behaviour where the source is in a high luminosity level for a short duration ($\sim$ 2\% of total cycle time) followed
by a longer spell in the low state. The spike like behaviour is observed during the $\rho$ class activity in GRS 1915+105. However, $\kappa$ and $\lambda$ classes are also observed frequently where the time spent by the source in peak luminosity is roughly equal to the time spent in dip luminosity. 
To address both these issues, assumptions of the standard accretion disc theory needs 
to be modified. Observed order of magnitude change in luminosity as well as $\kappa$/$\lambda$ class like light curves are successfully reproduced in a numerical simulation by \citet{b21}. They modified the viscosity prescription by changing the standard assumption that the viscous stress scales with the total pressure and allow for a fraction of the energy to be dissipated in the
corona above and a jet carrying a luminosity-dependent fraction of energy, to qualitatively explain the observations. However, \citet{b31} and \citet{b58} show that only the latter
maybe sufficient. More recently \citet{b60,b61} have studied the phase resolved spectra of the
`heartbeat' oscillation and show it to be consistent with the radiation pressure instability.

While a qualitative picture arises by these investigations, a complete spectral evolution
understanding of different oscillations is still not clear. Importantly, a model
independent fundamental question regarding the oscillations remains unexamined i.e., whether
the change in flux is due to the appearance of an additional spectral component or does the
total spectral shape evolve from the low to the high state? The question can be put in better context
when one notes that in other typical black hole systems, the prominent spectral states $-$ `hard'
and `soft' have completely different spectral shapes and temporal properties. One cannot explain
the difference between the two as an addition or subtraction of a spectral component. However,
for the oscillation of GRS 1915+105, it is not clear whether the change in spectral shape and
flux is due to an addition of a component or not. If indeed the oscillation is due to an
additional component appearing and disappearing, its flux variation could be much larger than
than the factor of 10 difference between the peak and dip levels. In the framework of the radiation pressure instability model, this
could mean that the accretion disc flux variation may indeed be several orders of magnitude as seen in numerical
simulations.

A recent outburst (2011) in another transient black hole X-ray binary (BHXB) IGR J17091$-$3624 
shows regular, repetitive, large amplitude oscillations in its X-ray
light curve, similar to the $\rho$ cycles observed in GRS 1915+105
\citep{b6,b26,b54}. The outburst was first observed by the {\it IBIS}
detector on-board {\it INTEGRAL} satellite in the 15.0$-$120.0 keV
energy range \citep{b19}. Using detailed timing and spectral analysis,
\citet{b54} show that prior to the $\rho$ class activity in IGR J17091$-$3624,
the source passes through the hard/soft intermediate state
which is also observed in GRS 1915+105 \citep{b84}. However,
\citet{b54} found that these oscillations are spectrally harder than
GRS 1915+105 by a factor of 2. Using X-ray morphology and hardness,
few more types of variabilities have been found \citep{b83,b81}, some of which are
similar to GRS 1915+105. Hence IGR J17091$-$3624 provides us a unique opportunity to test models and hypothesis which are used to explain the large amplitude, quasi-periodic X-ray oscillations in GRS 1915+105.

In this work, we examine the flux resolved spectroscopy of large amplitude oscillations in GRS 19115+105 and
in the new source IGR J17091$-$3624. One of our motivation is to test the p-free disc model
and to see if it empirically describes the thermal spectral component better than the
standard disc one. Our primary aim is to investigate whether the oscillations
observed in the two sources can be described by the appearance and disappearance of a
single spectral component. We will do so by assuming that the spectrum at the
dip of the oscillation is persistent and we will subtract it from the spectrum at the peak by treating it as the background spectra of the peak spectra in the {\tt XSpec}. We
will test if such a `difference' spectrum can be fitted by a simple single component. We also justify the difference spectral fitting method by another method in which we jointly fit peak and dip spectrum using a steady emission which is tied for both and a variable component which is free to vary. Section 2 and its subsections provide details
of the observations and analysis procedure. We discuss results in Section
3, while in section 4  we summarize the results and discuss their implications.

\section{Observation \& Data analysis}

In order to search observations which show large amplitude, quasi-regular X-ray variations similar to the $\kappa$/$\lambda$ class, we  analyse all data as observed by the {\it RXTE}/PCA between 1996 and 1999. We assume that results obtained from observations of these three years are valid for all similar observation in GRS 1915+105. To create difference spectra using flux resolved spectroscopy technique, we set two data selection criteria $-$
(i) time interval of any subsequent peak or dip of an oscillation must be greater than few multiples of 16 s in case of GRS 1915+105. The reason is full energy channel information (2.0$-$60.0 keV) is available only for {\tt Standard2} data in GRS 1915+105 which has 16 sec. time resolution. We adopt different criteria for observations in IGR J17091$-$3624. We consider here {\tt GoodXenon} data file which provide channel-energy information covering entire {\it RXTE}/PCA energy range with the time resolution of 10 $\mu$s. Since time-scale of quasi regular and large amplitude oscillations in IGR J17091$-$3624 is moderately smaller ($\sim$ 10$-$ 20 times) than those observed in $\kappa$ and $\lambda$ classes in GRS 1915+105, we use 0.25 s bintime to create peak and dip spectra, separately. We find that other choices of bin-size in extracting spectra results in poor signal-to-noise ratio due to low count rate and faster oscillation period. In order to improve the signal$-$to$-$noise ratio quality further, we merge several peak and dip intervals to carry out further analysis. In IGR JI7091$-$3624, the $\rho$ class activities are frequently observed with higher frequency than GRS 1915+105. For both sources, we do not use $\rho$ class observations in this work as its oscillation frequency is very high. Due to high frequency, changes in spectral parameters are very fast and the source spends few seconds at the peak flux level ($\sim$ 1/10 of dip time interval). Hence activities of both sources during the $\rho$ class do not fulfil our selection criteria. We consider observations which show light curves similar to the $\kappa$/$\lambda$ class in GRS 1915+105, where the dip time intervals are roughly similar to the following peak time intervals. 
(ii) The light curve should contain sufficiently large number of such oscillations so that several peak time intervals as well as dip time intervals can be merged to obtain sufficiently high signal$-$to$-$noise ratio for further analysis.

\begin{table*}
 \centering
 \caption{Observation details of $\lambda$ and $\kappa$ classes in GRS 1915+105 and corresponding peak time and dip time intervals used in our analysis}
\begin{center}
\scalebox{0.90}{%
\begin{tabular}{ccccccc}
\hline 
Obs-no & Obs-ID  & Date(dd-mm-yy) & MJD & Class & Total peak time(s) & Total dip time(s) \\
\hline 
1  & 10408-01-38-00 (1) & 07-10-96 & 50363.27 & $\lambda$ & 1801.2 & 1002.5 \\
2  & 10408-01-38-00 (2) & 07-10-96 & 50363.34 & $\lambda$ & 1440.5 & 925.7 \\
3  & 10408-01-38-00 (3) & 07-10-96 & 50363.41 & $\lambda$ & 1500.3 & 860.4 \\
4  & 20186-03-01-01 (1) & 28-06-97 & 50627.55 & $\kappa$  & 1160.4 & 1161.5 \\
5  & 20186-03-01-01 (2) & 28-06-97 & 50627.62 & $\kappa$  & 1402.3 & 1117.6 \\
6  & 20402-01-36-00 (1) & 10-07-97 & 50639.63 & $\lambda$ & 1008.3 & 1193.6 \\
7  & 20402-01-36-00 (2) & 10-07-97 & 50639.74 & $\lambda$ & 1033.8 & 950.5 \\
8  & 20402-01-37-01 (1) & 12-07-97 & 50641.49 & $\lambda$ & 1218.3 & 1330.7 \\
9  & 20402-01-37-01 (2) & 12-07-97 & 50641.56 & $\lambda$ & 1311.5 & 969.8 \\
10 & 40403-01-06-02     & 16-04-99 & 51284.16 & $\kappa$  & 494.8  & 474.3 \\
11 & 40403-01-06-00    & 16-04-99 & 51284.09 & $\kappa$  & 692.5  & 772.2 \\
12 & 40403-01-06-000 (1)& 16-04-99 & 51284.28 & $\kappa$ & 1120.7 & 899.5 \\
13 & 40403-01-06-000 (2)& 16-04-99 & 51284.35 & $\kappa$ & 936.4 & 899.5 \\
14 & 40403-01-06-000 (3)& 16-04-99 & 51284.41 & $\kappa$ & 775.7 & 752.3 \\
15 & 40403-01-06-000 (4)& 16-04-99 & 51284.49 & $\kappa$ & 1077.3 & 752.7 \\
16 & 20402-01-33-00 (2) & 18-06-97 & 50617.61 & $\kappa$ & 896.3  & 864.5 \\
17 & 20402-01-33-00 (1) & 18-06-97 & 50617.54 & $\kappa$ & 1952.6 & 1696.2 \\
\hline
\end{tabular}}
\end{center}
\end{table*}

Out of the selected three years observation of the {\it RXTE}/PCA, we find 17 observations in GRS 1915+105 which fulfil our selection criteria. We also exclude $\beta$ and $\alpha$ class observations from our analysis as they lack stable peak flux level. The details of the selected 17 observations of $\kappa$ and $\lambda$ classes are provided in the Table 1. Using 1 s bin$-$time, we extract light curve of each observation in the energy range 2.0$-$60.0 keV. For each quasi-regular oscillation, we define the dip time interval and peak time interval as time intervals of the persistent flux level/low stable flux level and peak flux level/high stable flux level between sharp rise and sharp fall respectively. To select peak and dip intervals, we consider both X-ray intensity profile and hardness ratio profile. We observe that within consecutive burst-to-dip and dip-to-burst transitions in the X-ray intensity profile, both peak and dip intervals are clearly demarcated by sharp changes in the time-dependent hardness ratio profile. Using this characteristic as the peak and dip selection criteria, we approximately set threshold values in X-ray intensity to choose both peak and dip intervals which are consistent with the sharp change in the hardness ratio. Such selections are shown by green and red colours in all panels of Figure 1 and threshold values are shown by horizontal dotted lines in top four panels of Figure 1. We follow the similar criteria for selecting peak and dip intervals in IGR J17091$-$3624.
 
We produce the hardness intensity diagram (HID) and the colour$-$colour diagram (CD) using binned light curves in different energy bands. Hard colour is defined as the ratio of the background subtracted count rate between 12.0$-$60.0 keV and 2.0$-$5.0 keV and soft colour is defined as the ratio of the background subtracted count rate between 5.0$-$12.0 keV and 2.0$-$5.0 keV. To study characteristics of the source during peak flux level and persistent flux level, we separately produce unfolded rms normalized power density spectra (PDS; in the unit of rms$^2$) in the 0.01$-$10.0 Hz frequency range. We use 1/64 s bin$-$time so that the Nyquist frequency would be 32 Hz. We extract energy spectra and background spectra for peak flux level and persistent flux level using {\tt Standard2} data file with 16 s time resolution.

\begin{table*}
 \centering
 \caption{Average spectral parameter values for all observations of peak and dip spectra when fitted with {\tt diskpbb+nthcomp} model in GRS 1915+105 and IGR J17091$-$3624. Range of spectral parameters for all observations (i.e., maximum and minimum values of the corresponding parameter observed from all observations) are given in braces. The unit of fluxes is 10$^{-8}$ erg$-$s$^{-1}$$-$cm$^{-2}$ and the unit of luminosity is 10$^{38}$ erg$-$s$^{-1}$. Epoch of the corresponding observation is provided in superscript with $\chi^2$/d.o.f .}
\begin{center}
\scalebox{0.75}{%
\begin{tabular}{ccccccccccccccc}
\hline
spectra & kT$_{in}$ & R$_{in}$ & `{\it p}' parameter & Powerlaw photon & kT$_{e}$ &  L$_{total}$ & F$_{total}$ & F$_{diskpbb}$ & F$_{nthcomp}$ &  $\chi^2/d.o.f$ \\
intervals     & (keV) & (km) & value & index ($\Gamma$) & (keV) &  & & &  &    \\  
\hline
{\bf\large GRS 1915+105} & & & & & & & & &  & & \\
\hline
Peak spectra & $2.37^{+0.04}_{-0.04}$ & $32.45^{+4.15}_{-3.23}$ & 0.525$^{+0.001}_{-0.002}$ & $2.55^{+0.05}_{-0.03}$ & 6.10$^{+1.56}_{-2.01}$ &  13.92 & 8.09$^{+1.06}_{-0.63}$ & 6.19$^{+1.14}_{-1.23}$ & 1.85$^{+0.82}_{-0.57}$ & 1.09(40$^{\bf 3}$,35$^{\bf 4}$) \\
(Range) & (2.27$-$2.49) & (24.13$-$41.56) & (0.501$-$0.541) & (2.35$-$2.69) & (4.16$-$6.95) &  (9.77$-$19.86) & (7.39$-$11.49) & (4.88$-$7.53) & (1.37$-$4.39) & (0.98$-$1.18) \\
\hline
Dip spectra & $1.29^{+0.05}_{-0.03}$ & $48.37^{+3.78}_{-3.29}$ & 0.643$^{+0.032}_{-0.015}$ & $1.77^{+0.04}_{-0.05}$ & 9.69$^{+2.55}_{-2.08}$ &  2.05 & 1.19$^{+0.22}_{-0.35}$ & 0.27$^{+0.13}_{-0.08}$ & 0.89$^{+0.12}_{-0.11}$ & 1.07(40$^{\bf 3}$,35$^{\bf 4}$) \\
(Range) & (1.24$-$1.35) & (41.74$-$53.49) & (0.599$-$0.692) & (1.61$-$1.89) & (5.34$-$11.57) & (1.76$-$2.34) & (0.95$-$1.34) & (0.14$-$0.41) & (0.68$-$0.99) & (1.04$-$1.16) \\
\hline
\hline
{\bf\large IGR J17091$-$3624} & & & & & & &  & & & & \\
\hline
Peak spectra & $1.94^{+0.21}_{-0.18}$ & -- & 0.515$^{+0.024}_{-0.011}$ & $1.38^{+0.78}_{-0.11}$ & 7.35$^{+3.59}_{-1.06}$ &  -- & 0.21$^{+0.03}_{-0.03}$ & 0.17$^{+0.01}_{-0.03}$ & 0.04$^{+0.01}_{-0.01}$ & $42.57/39^{\bf 5}$ \\
(Range) & (1.81-2.21) & -- & (0.505-0.531) & (1.25-2.18) & (6.12-15.89) &  -- & (0.18-0.26) & (0.14-0.22) & (0.02-0.05) & (37.85-46.34) \\
\hline
Dip spectra & $0.42^{+0.16}_{-0.09}$ & -- & 0.714$^{+0.013}_{-0.045}$ & $2.18^{+0.25}_{-0.95}$ & 21.59$^{+4.41}_{-3.96}$ &  -- & 0.04$^{+0.02}_{-0.02}$ & 0.01$^{+0.01}_{-0.03}$ & 0.03$^{+0.03}_{-0.01}$ & $44.62/39^{\bf 5}$ \\
(Range) & (0.31$-$0.77) & -- & (0.668$-$0.739) & (1.18$-$2.45) & (15.85$-$25.69) &  -- & (0.02$-$0.07) & (0.01$-$0.03) & (0.03$-$0.04) & (40.26$-$48.23) \\
\hline
\end{tabular}}
\end{center}
\end{table*}

\begin{figure*}
\begin{center}
\begin{tabular}{c|c}
\includegraphics[scale=0.30,angle=-90]{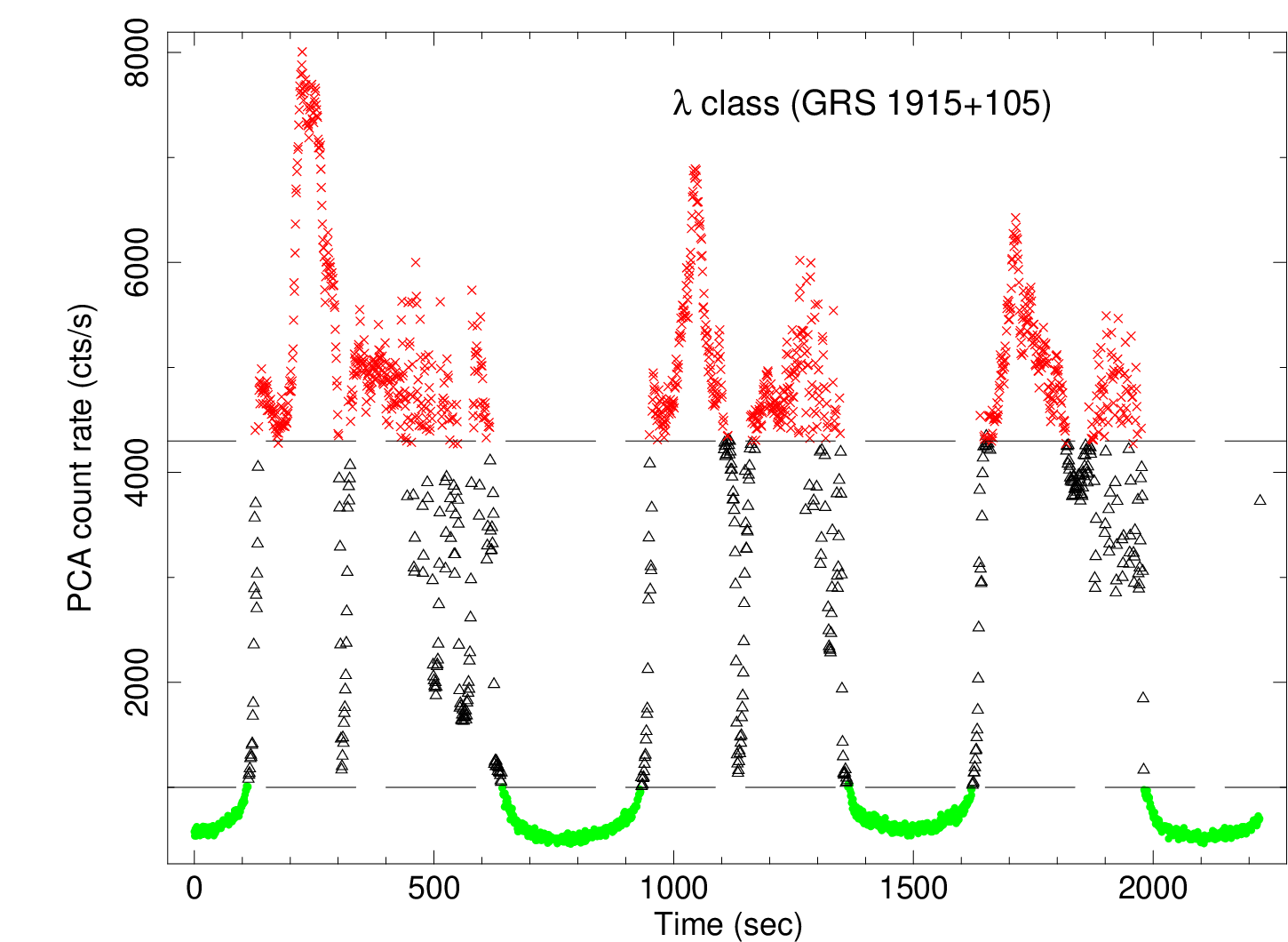} &
\includegraphics[scale=0.30,angle=-90]{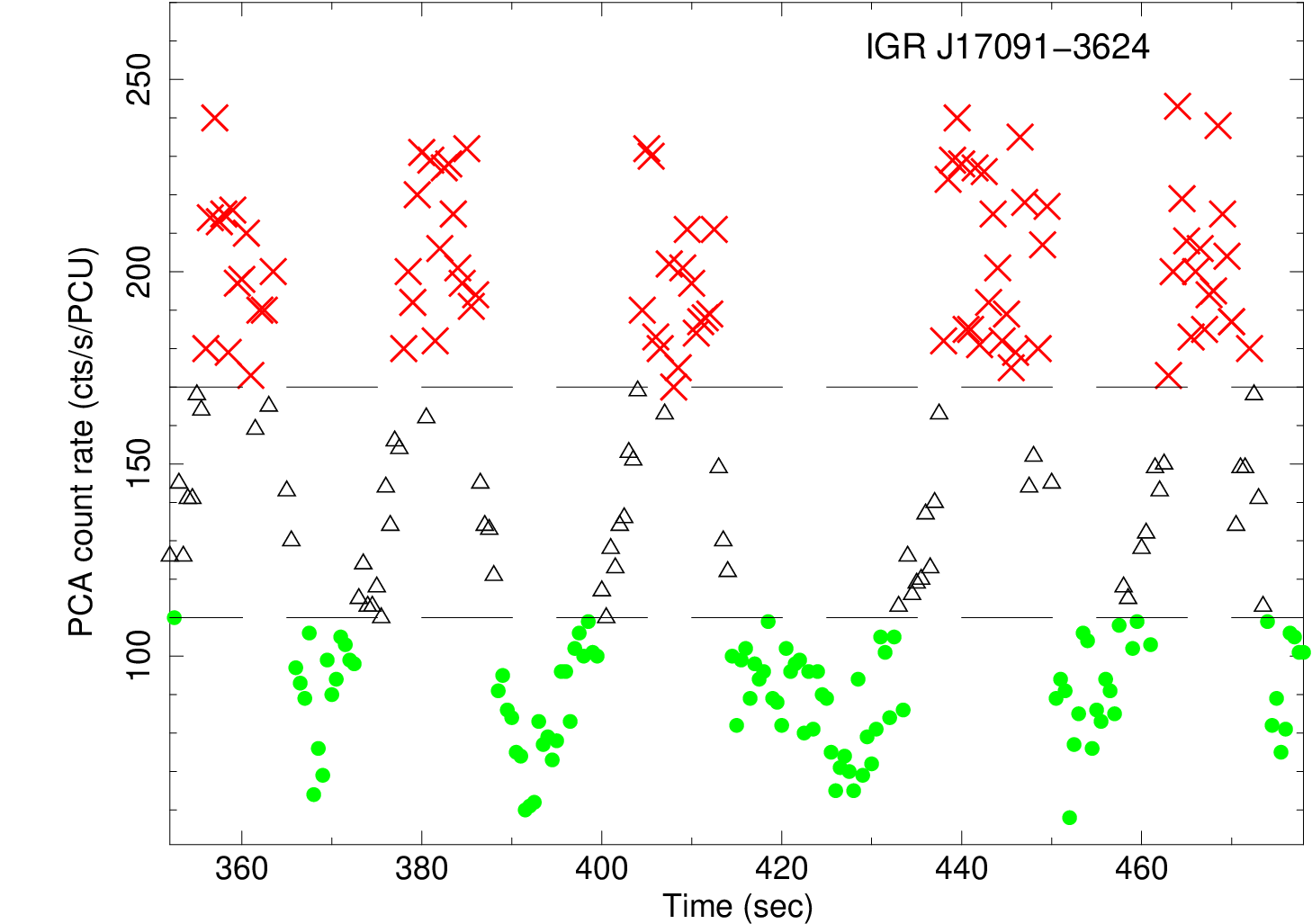} \\
\includegraphics[scale=0.30,angle=-90]{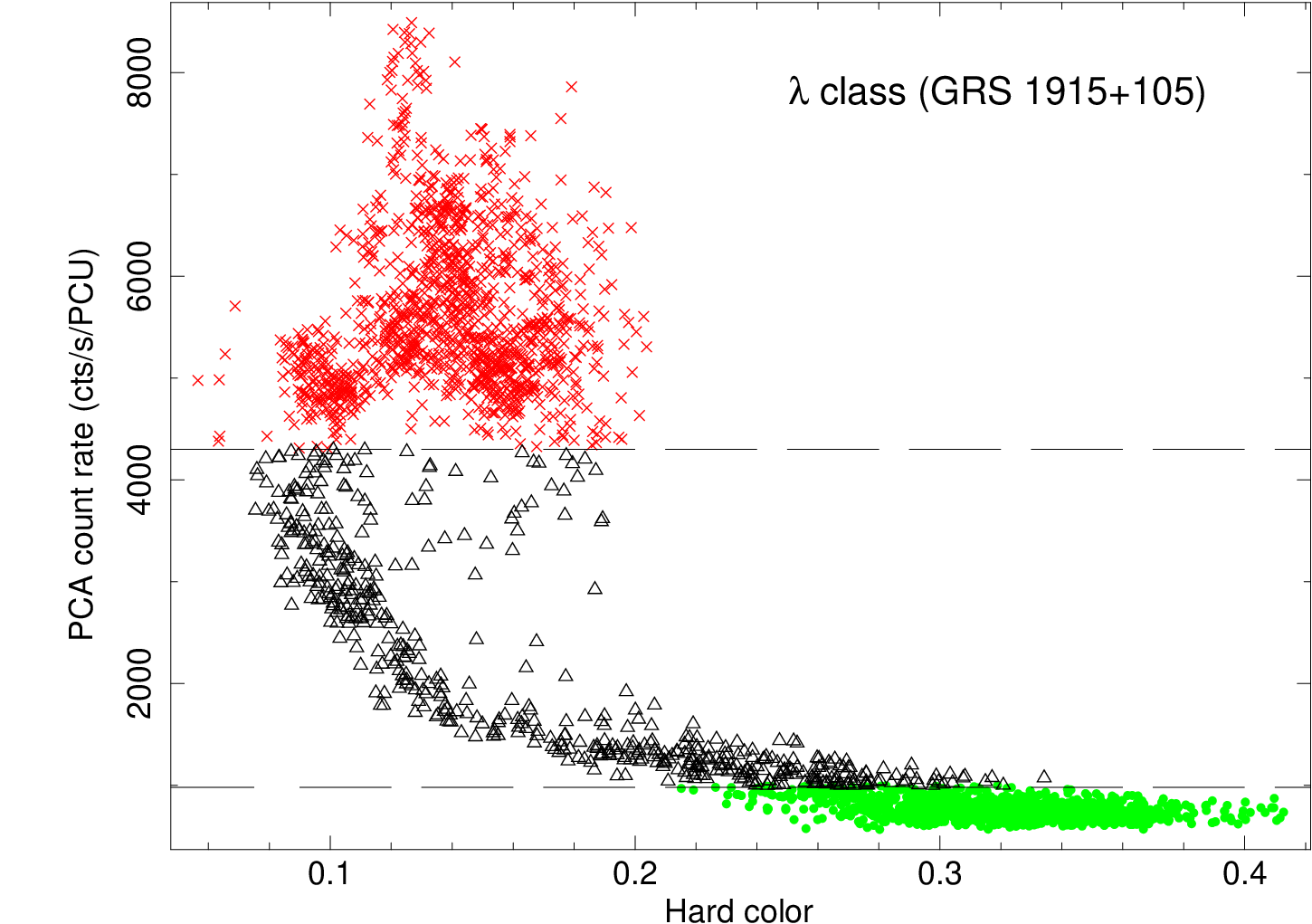} &
\includegraphics[scale=0.30,angle=-90]{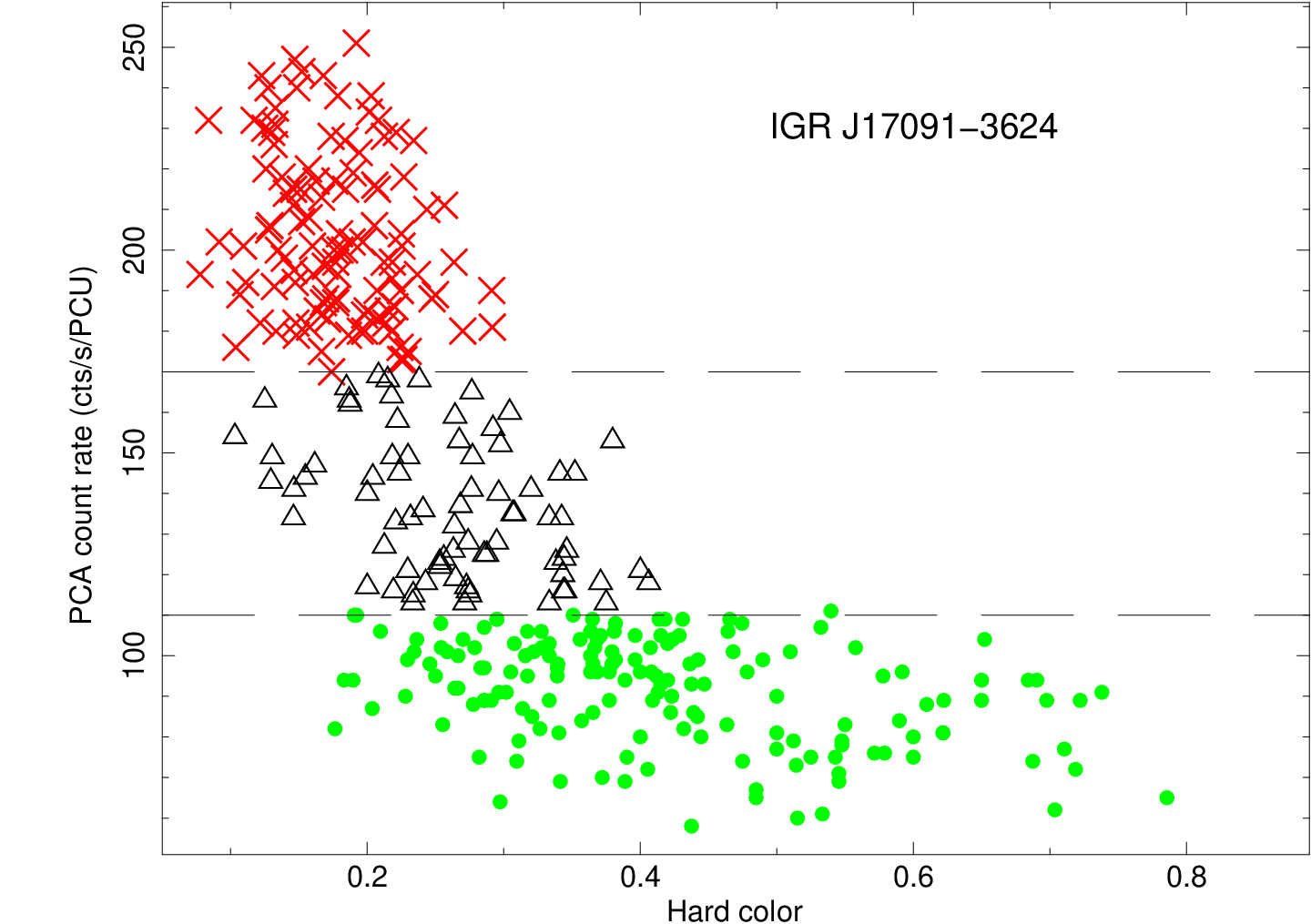} \\
\includegraphics[scale=0.30,angle=-90]{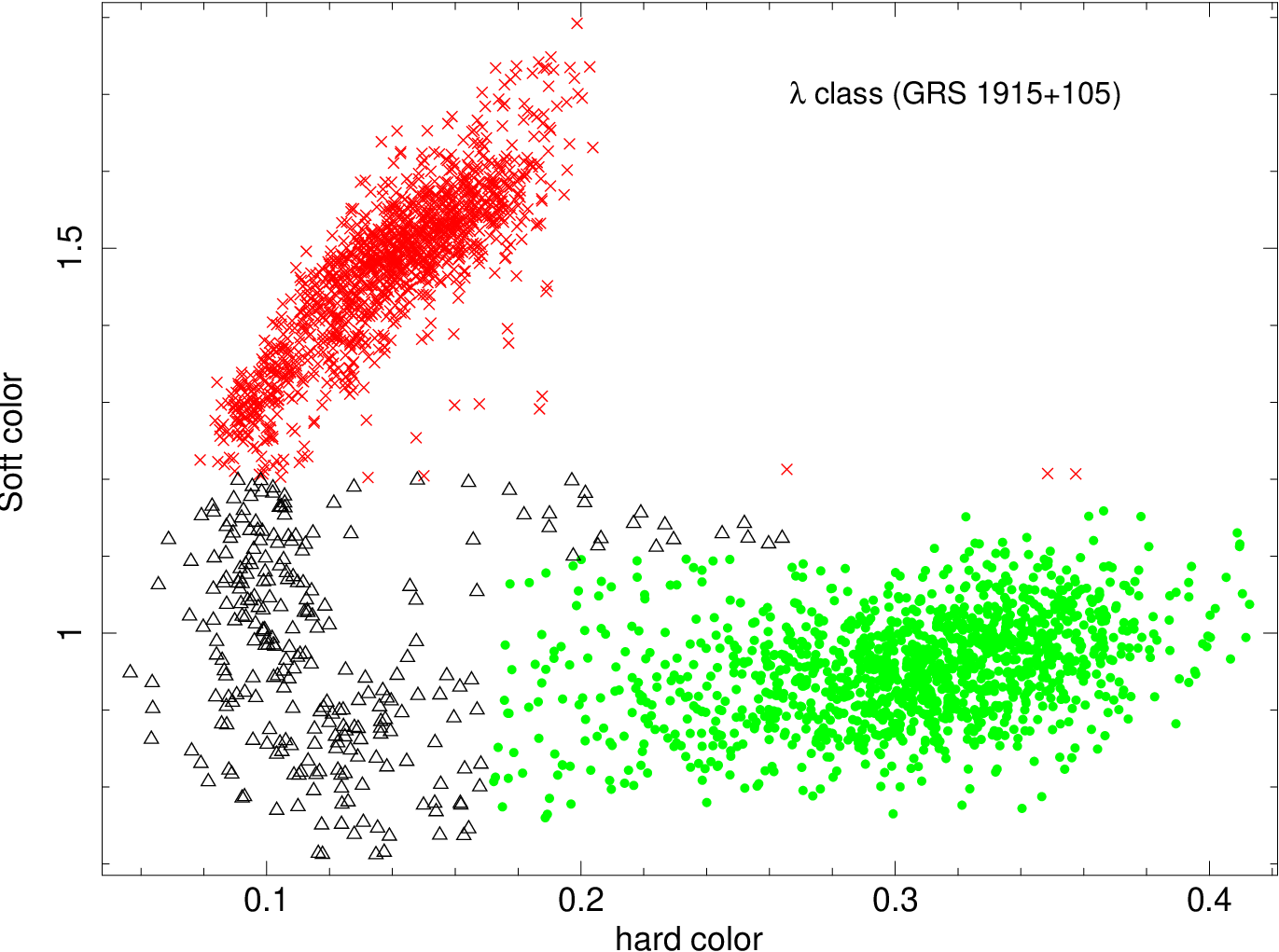} &
\includegraphics[scale=0.30,angle=-90]{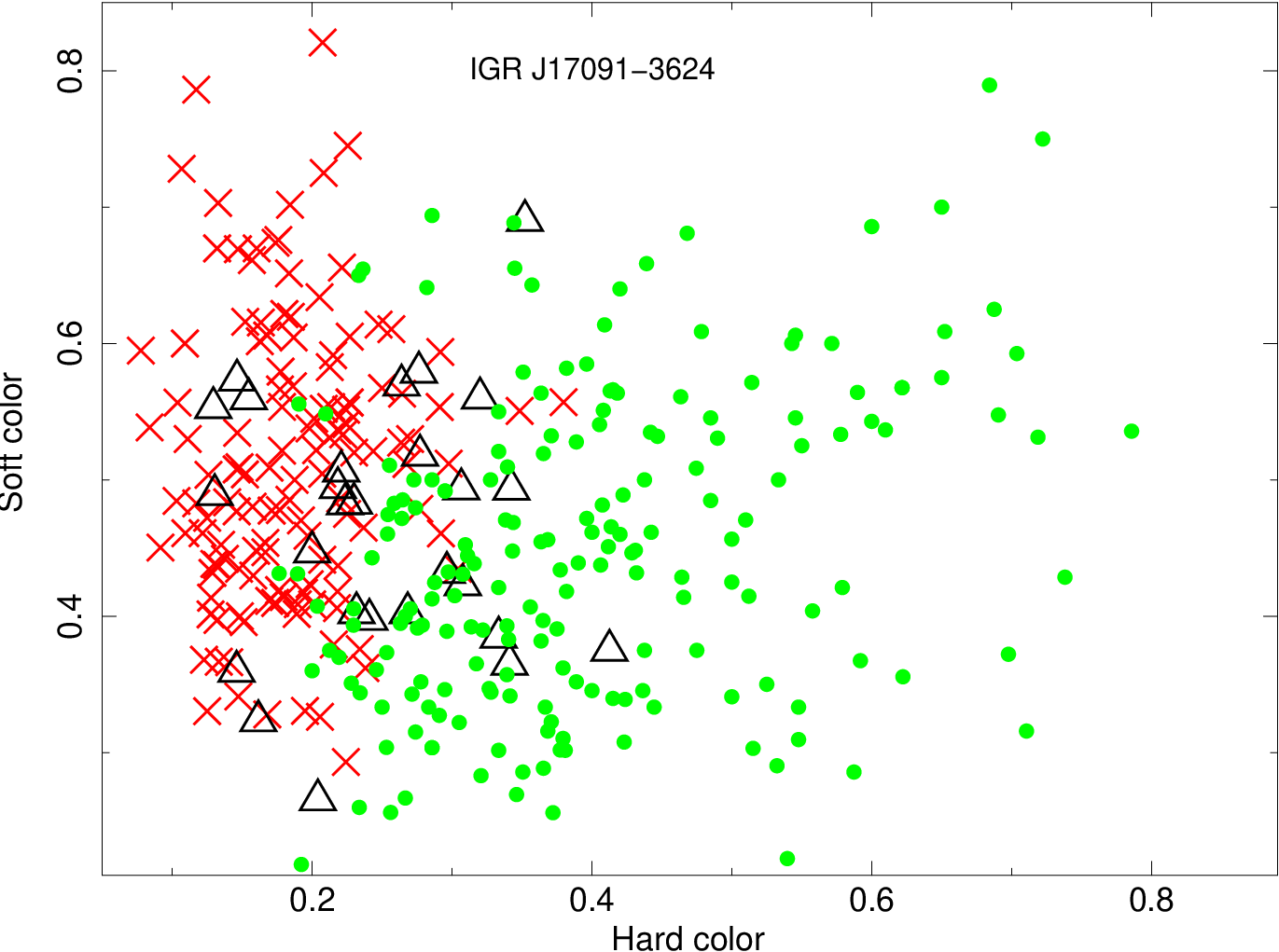} \\
\end{tabular}
\caption{{\it Top panels}: 1 s binned 2.0$-$60.0 keV light curve of a $\lambda$ class observation in GRS 1915+105 (top left panel) and the light curve of similar observation in IGR J17091$-$3624 (top right panel). The selection in light curve for extracting peak spectrum is shown in red and the selection in light curve for extracting dip spectrum is shown in green. {\it Middle panels:} the HID and the CD of the $\lambda$ class in GRS 1915+105 (middle left and bottom left panel) and IGR J17091$-$3624 (middle right and bottom right panel) are shown. Red, black and green colours in the HID and CD represent time intervals shown in red, black and green colour respectively in corresponding light curves (top panels). In top four panels, threshold values for selecting peak and dip intervals are shown by horizontal dotted line.}
\end{center}
\end{figure*}

\subsection{Spectral analysis}

Since the fits to {\it RXTE}/PCA spectra of the Crab nebula reveal residuals as large as 1\%, and we therefore add the systematic error of 1\% to all PCU energy channels using {\tt grppha} \citep{b76} tool in the {\tt HEASOFT 6.12}. Below 3 keV, a large residual is often observed which cannot be associated with any plausible spectral feature and above 22 keV, the difference spectrum suffers from large background counts and the calibration uncertainty. Hence to maintain uniformity, we restrict our {\it RXTE}/PCA spectral analysis to the 3.0$-$22.0 keV band for peak, dip and difference spectra analysis. For difference spectra analysis of IGR J17091$-$3624, we use 3.0$-$18.0 keV energy range due to high background counts above 18.0 keV. All PCA spectra are dead-time corrected following the procedure given by \citet{b76}. While fitting peak, dip and difference spectra of GRS 1915+105, we fix the value of the hydrogen column density at 5.0 $\times$ 10$^{22}$ cm$^{-2}$. This is consistent with the {\it Chandra}/HETGS and {\it RXTE}/PCA joint analysis by \citet{b72} and lies between $\sim$4.0 $\times$ 10$^{22}$ cm$^{-2}$ [determined from an analysis of the {\it ASCA}/GIS data and {\it RXTE}/PCA data for GRS 1915+105 by \citet{b76,b32}] and $\sim$5.6 $\times$ 10$^{22}$ cm$^{-2}$ [determined from the analysis of {\it BeppoSAX} data by \citet{b33}]. In case of IGR J17091$-$3624, we fix it at 0.9 $\times$ 10$^{22}$ cm$^{-2}$ \citep{b53}.

\subsubsection{Choices of model components}  

An important step in this work is the choice of spectral models to describe peak, dip and difference spectra. We start with single component models like {\tt diskbb}, {\tt powerlaw}, {\tt nthcomp} and {\tt diskpbb}. While fitting peak spectra and dip spectra, we find that single component models provide unacceptably large reduced $\chi^2$. In order to improve the fit (i.e, reduced $\chi^2$), we use two-component models like {\tt diskbb+powerlaw}, {\tt diskbb+nthcomp}, {\tt powerlaw+nthcomp} and {\tt diskpbb+nthcomp}. 

\begin{figure*}
  \begin{center}
  \begin{tabular}{c|c}
\includegraphics[scale=0.31,angle=-90]{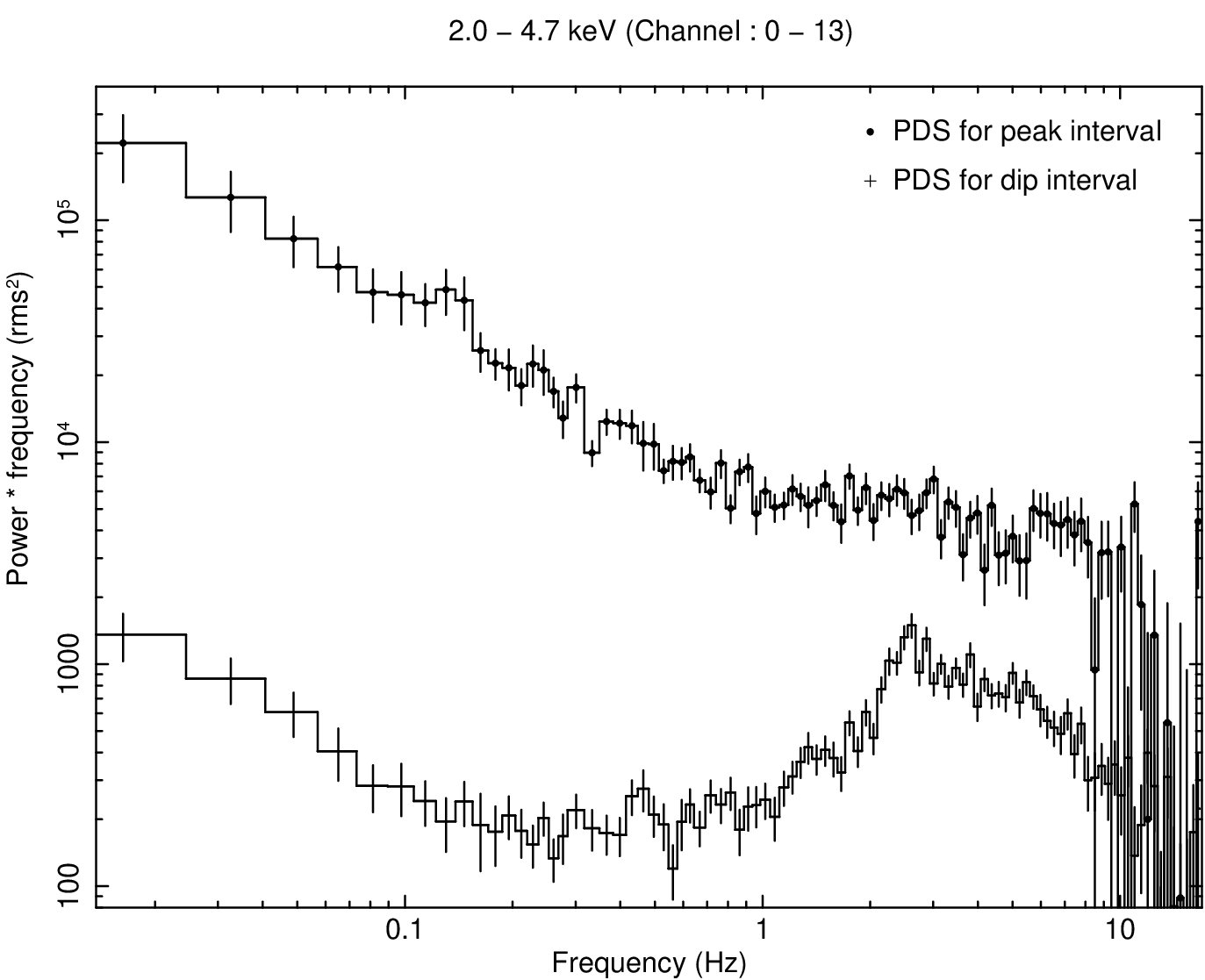} &
\includegraphics[scale=0.31,angle=-90]{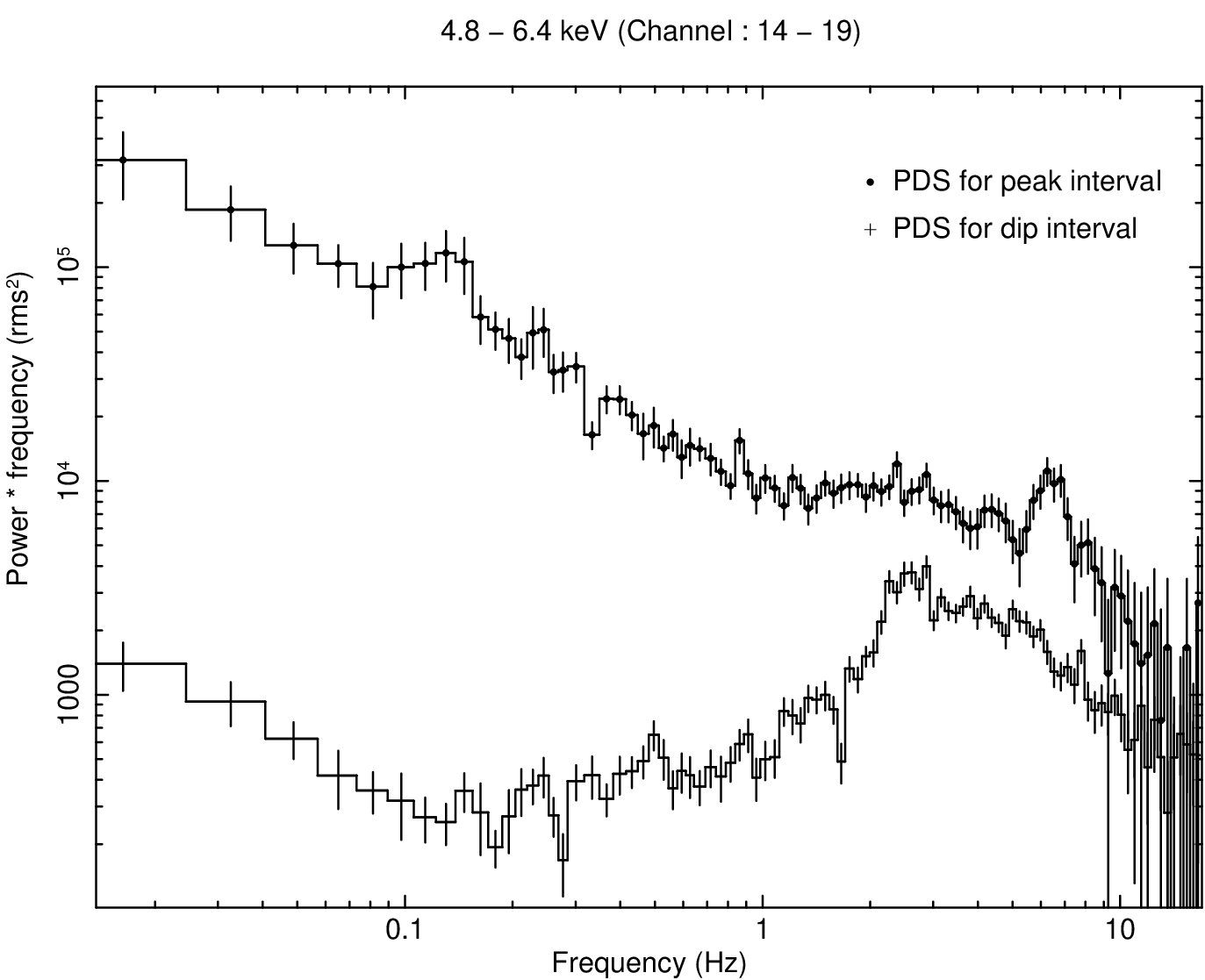} \\
\includegraphics[scale=0.31,angle=-90]{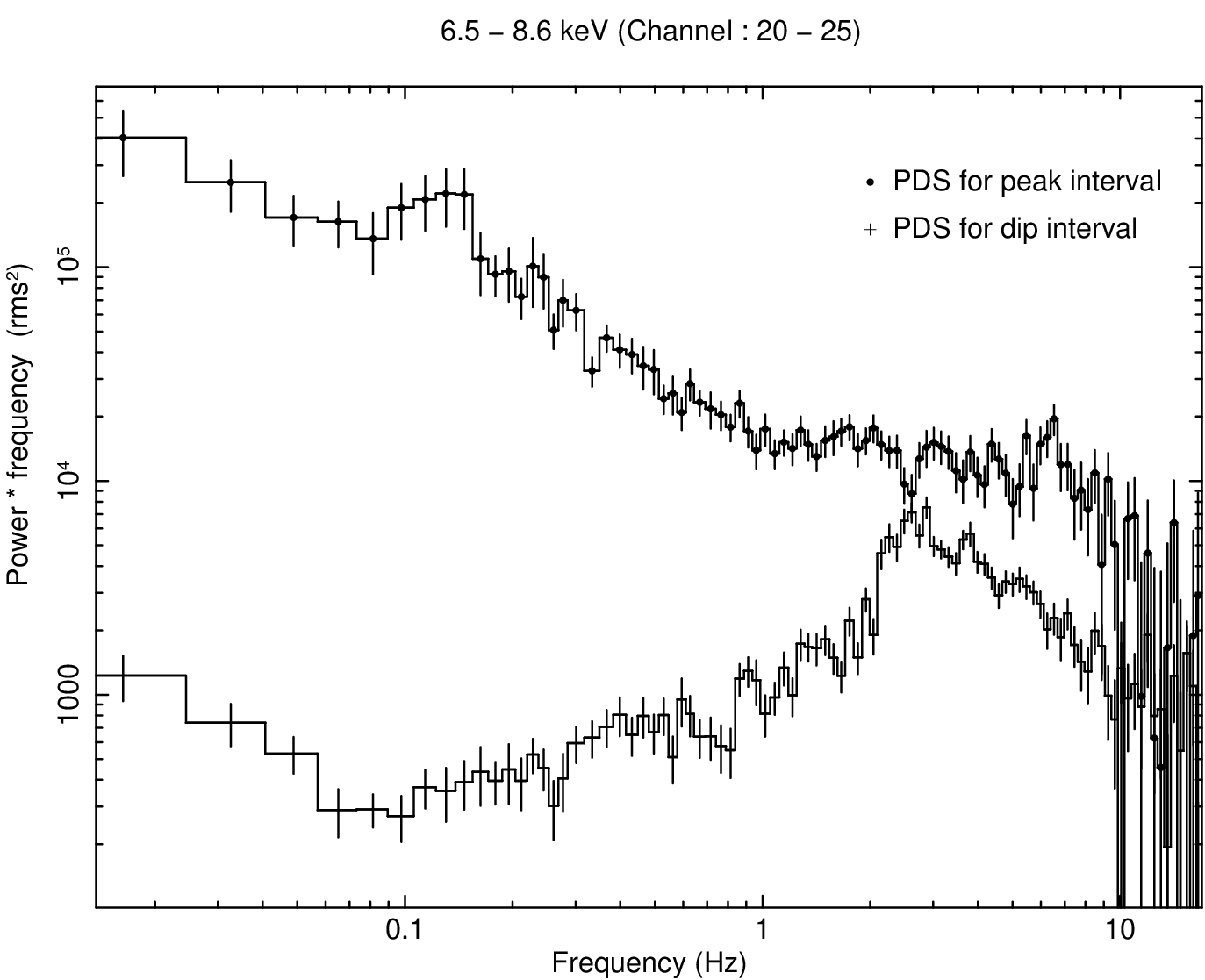} &
\includegraphics[scale=0.31,angle=-90]{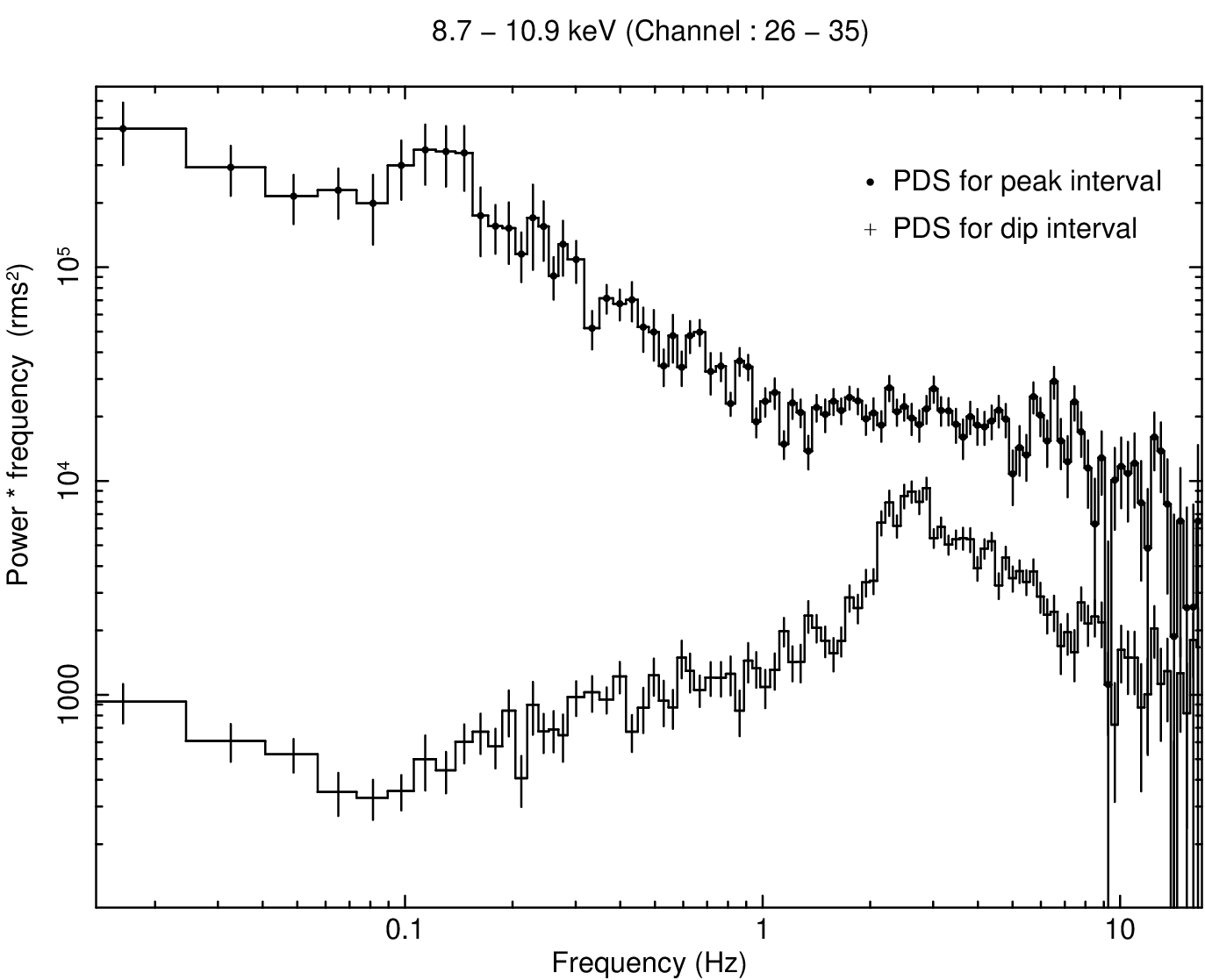} \\
\end{tabular}
\caption{rms normalized PDS during peak and dip intervals for different energy bands : 2.0$-$4.7 keV (top$-$left panel), 4.8$-$6.4 keV (top$-$right panel), 6.5$-$8.6 keV (bottom$-$left panel) and 8.7$-$10.9 keV (bottom$-$right panel). In all panels, PDS during peak time intervals always show excess power than PDS during dip time intervals at all frequencies.}
\end{center}
\end{figure*}

\begin{table*}
 \centering
 \caption{Average reduced $\chi^2$ values of fitting simultaneous peak and dip spectra (for epoch 3 only) using different model combinations in GRS 1915+105}
\begin{center}
\scalebox{0.95}{%
\begin{tabular}{|llllllllll|}
\hline 
Variable &\vline & Steady & Emission & Component & \\
\cline{3-6}
component &\vline & {\tt diskbb+nthcomp} & {\tt diskbb+powerlaw} & {\tt diskpbb+powerlaw} & {\tt diskpbb+nthcomp} \\ 
\hline
\hline 
{\tt diskbb} &\vline & 2112.45/86 & 2244.89/88 & 945.27/87 & 717.53/85 \\
{\tt diskpbb} &\vline & 137.15/85 & 109.52/87 & 122.34/86 & 88.56/84 \\
{\tt nthcomp} &\vline & 145.69/84 & 391.39/86 & 434.73/85 & 128.95/83 \\
{\tt powerlaw} &\vline & 11206.57/86 & 17722.48/88 & 8781.35/87 & 7872.76/85 \\
\hline
\end{tabular}}
\end{center}
\end{table*}

We also add a Gaussian model component at 6.3 keV. We find that among all two component model fitting, the combination of {\it p}$-$free multi temperature disc blackbody ({\tt diskpbb} in {\tt XSpec}) and a thermal Comptonization model ({\tt nthcomp} in {\tt XSpec}) can adequately describe both peak and dip spectra for all observations. The innermost temperature of seed photons input to the {\tt nthcomp} model and that of the MCD component ({\tt diskbb} and {\tt diskpbb}) are tied to each other. We use disc blackbody option as the input type of seed photons in the {\tt nthcomp} model.
While fitting the spectra due to peak intervals with {\tt diskbb}, {\tt powerlaw}, {\tt nthcomp}, {\tt diskbb+powerlaw}, {\tt diskbb+nthcomp} and {\tt diskpbb+nthcomp} models, the average reduced $\chi^2$ of the fitting of epoch 3 observations (for example) are 3791.35/46, 7872.41/46, 738.37/44, 107.95/44, 68.24/41 and 43.61/40 respectively. While fitting the spectra due to dip intervals with earlier models, the average reduced $\chi^2$ of the fitting of epoch 3 observations (for example) are 2067.37/46, 375.48/46, 738.65/44, 167.77/44, 152.39/41 and 42.80/40 respectively. From both peak and dip spectral fitting with two component models, it may be noted that along with thermal Comptonization, a {\it p}$-$free disc model fits significantly better than the standard disc blackbody model with the significance of 5.1$\sigma$ and 6.5$\sigma$ respectively. Spectral parameters of the fitted peak and dip spectra using the {\tt diskpbb+nthcomp} model, averaged over all observations from both epoch 3 and epoch 4, are provided in Table 2 for both GRS 1915+105 and IGR J17091$-$3624. Including error measurements, range of spectral parameters obtained from all observations are provided in braces.   

In order to determine the spectroscopic nature of the large amplitude oscillations in GRS 1915+105 and IGR J17091$-$3624, which is the prime focus in this work, we use two new methods for spectral analysis. 

\subsubsection{Difference spectral fitting method}

Difference spectrum is defined as the difference of the spectral energy density between the spectrum due to peak intervals and the spectrum due to dip intervals within the same class. However, in stead of subtracting dip spectral count rate from peak spectral count rate, we use peak spectra as the source spectra in the {\tt XSpec v12.7.1} and we use the dip spectra as the background spectra of the peak spectra in the {\tt XSpec v12.7.1}. There are two advantages of using this method $-$ (1) as peak and dip time intervals have different exposure time, exposure time corrections are needed to consider. While subtracting background from the source spectra, the {\tt XSpec} take care of exposure time corrections. (2) The subtraction in {\tt XSpec} is corrected for background and area scaling values which represent background flux corresponds to the same area as the observation. Hence, for the {\it I}$_{th}$ channel, the difference spectra/dip spectra-subtracted count rate is given by \citep{b97} -

\begin{equation}
S_{diff}(I) = \frac{S_{peak}(I)}{A_{peak}(I)*T_{peak}} - \frac{B_{peak}(I)*S_{dip}(I)}{B_{dip}(I)*A_{dip}(I)*T_{dip}}
\end{equation}

where S$_{peak}$(I), S$_{dip}$(I) are spectral count rate during peak time and dip time intervals, T$_{peak}$ and T$_{dip}$ are exposure times of both intervals and B$_{peak}$(I), B$_{dip}$(I), A$_{peak}$(I) and A$_{dip}$(I) are background and area scaling values for both peak and dip intervals respectively.
 Now both peak spectrum and dip spectrum contain spectral counts from the source as well as from the background. We find that - (1) for both peak and dip intervals, background count rate is 1-2\% of source count rate and (2) at each individual energy bins, the difference between background spectral count during peak and dip time intervals is less than 1\% of the difference between peak and dip source spectral count rate. Since the background count rate difference is very low, any difference in background spectral shape between peak and dip spectra do not affect our analysis. Thus any excess emission over dip spectra which could give rise to the peak spectra can be realized using difference spectra method. However, in any case, care should be taken if peak and dip spectral count rate are comparable.  

Surprisingly, the difference spectra can be fitted equally well with the single component {\tt diskpbb} model or {\tt nthcomp} model. However, standard disc blackbody ({\tt diskbb}) or colour factor corrected disc blackbody ({\tt ezdiskbb}) model do not provide satisfactory fit (i.e., reduced $\chi^2$ is greater than 2 in both cases). We prefer to carry out our analysis with the {\tt diskpbb} model for two reasons : (1) In most cases, {\tt diskpbb} model provide better reduced $\chi^2$ values (0.9 $<$$\chi^2$/d.o.f $<$ 1.2) than {\tt nthcomp} ($<$0.8) and better constraint on error bars than the {\tt nthcomp} model. (2) It is simpler than {\tt nthcomp} since the former has one parameter less than the later. 
We find that the choice of single component {\tt diskpbb} model over {\tt nthcomp} for fitting difference spectra is also valid when we perform the energy spectra analysis of difference spectra in another BHXB IGR JI7091$-$3624. Difference spectra in IGR J17091$-$3624 can be fitted well with the single component {\tt diskpbb} model. In order to justify the credibility of this method, we use another method for a sanity check.  

\subsubsection{Joint spectra with a variable component fitting method}

In this method, we perform joint spectral fitting of peak and dip spectra. We fit spectra from both peak and dip time intervals with a model consisting of a steady emission plus a variable component. Parameters of the steady emission model are tied together for both time intervals. The variable component is kept free to vary in peak spectra and its normalization is kept at zero at dip spectra assuming it is not present in the dip spectra. For modelling steady emission, we use various two component models consisting of a disc emission and a powerlaw emission or Comptonization. For each combination of steady emission model and variable component, the average reduced $\chi^2$ of epoch 3 observations (for example) are given in the Table 3. It can be clearly noted that the combination of {\tt diskpbb} and {\tt nthcomp} model as the steady emission plus a {\tt diskpbb} model as a variable component provides significantly better reduced $\chi^2$ (at least 3$\sigma$ significance than its nearest value using {\it F}$-$test) than all other models. This result is remarkably consistent with the earlier method where we find that difference spectra can be fitted well with a single {\tt diskpbb} model. Table 4 \& 5 show details of the fitted parameter values of {\tt diskpbb} model along with 1$\sigma$ error-bars using both methods for GRS 1915+105 and IGR J17091$-$3624 respectively. 

\begin{figure*}
  \begin{center}
  \begin{tabular}{c|c}
\includegraphics[scale=0.30,angle=-90]{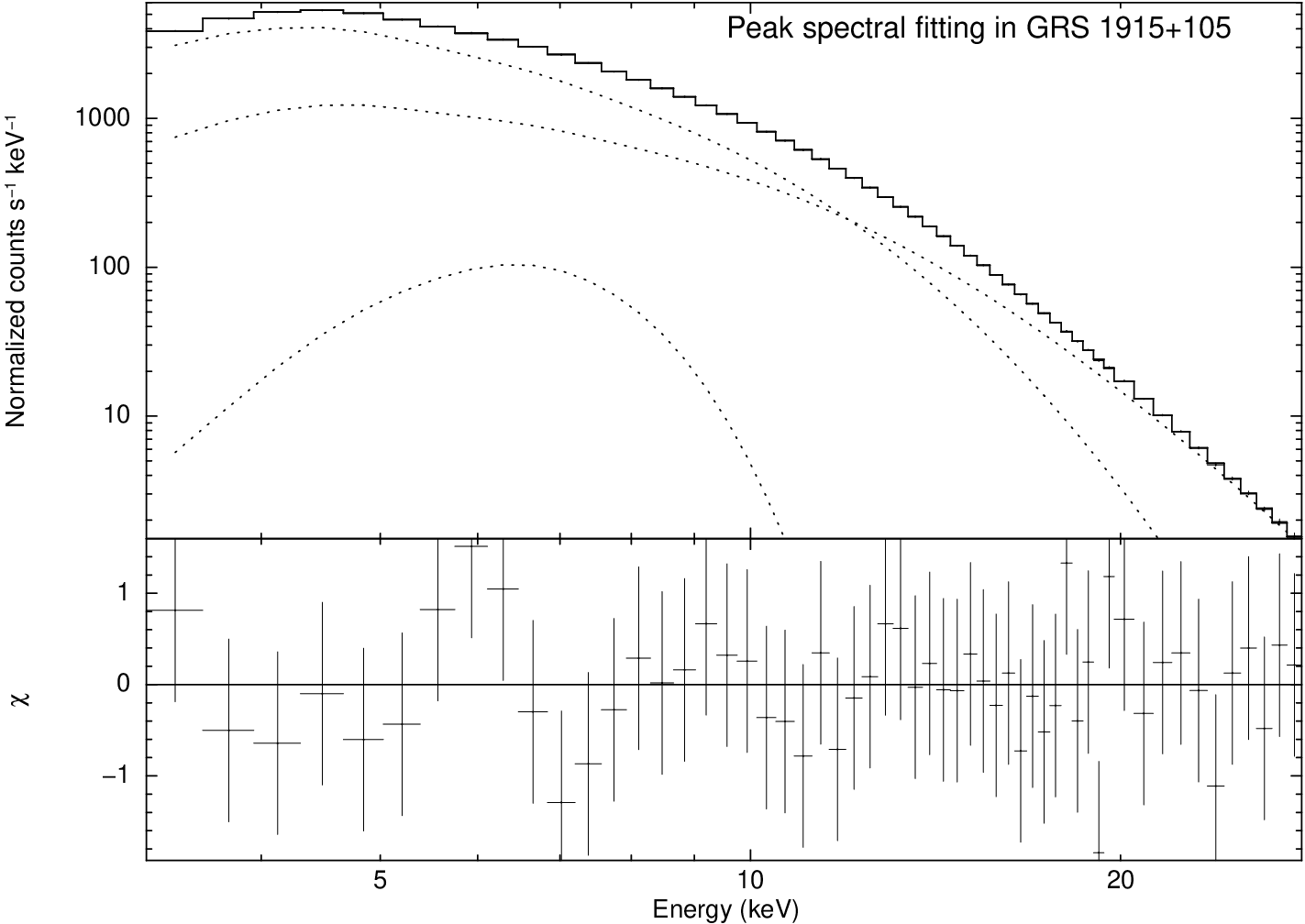} &
\includegraphics[scale=0.30,angle=-90]{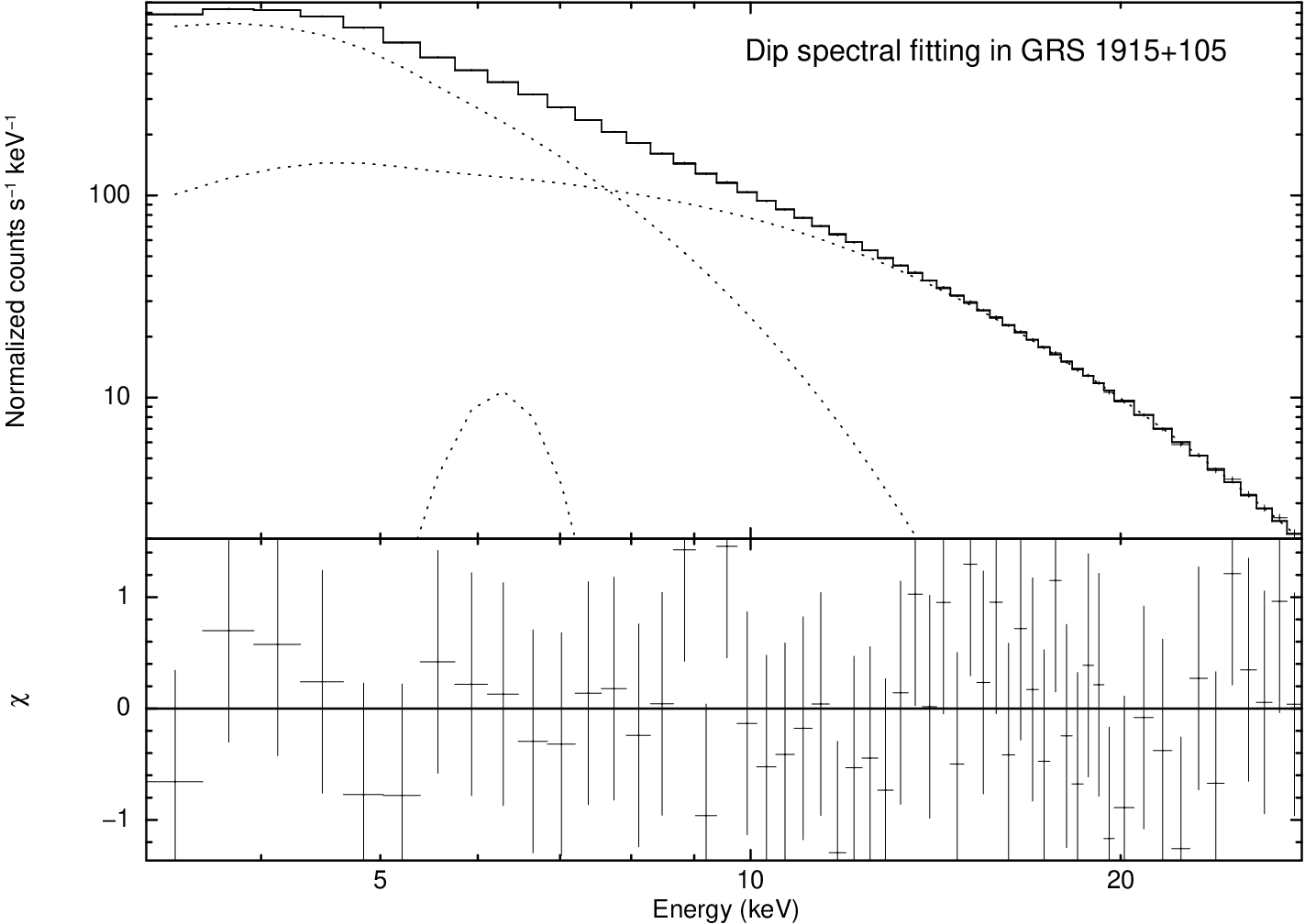} \\
\includegraphics[scale=0.30,angle=-90]{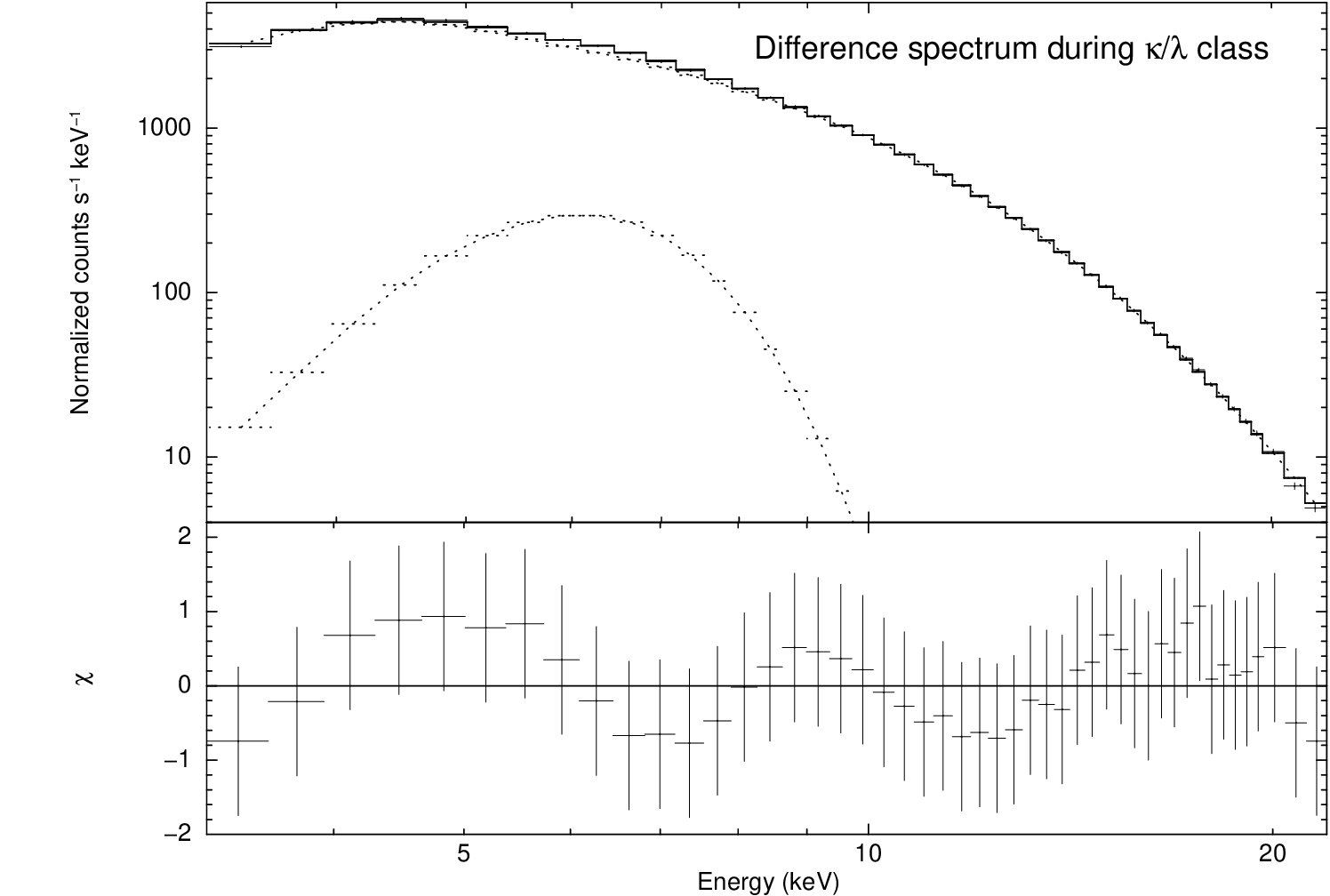} &
\includegraphics[scale=0.30,angle=-90]{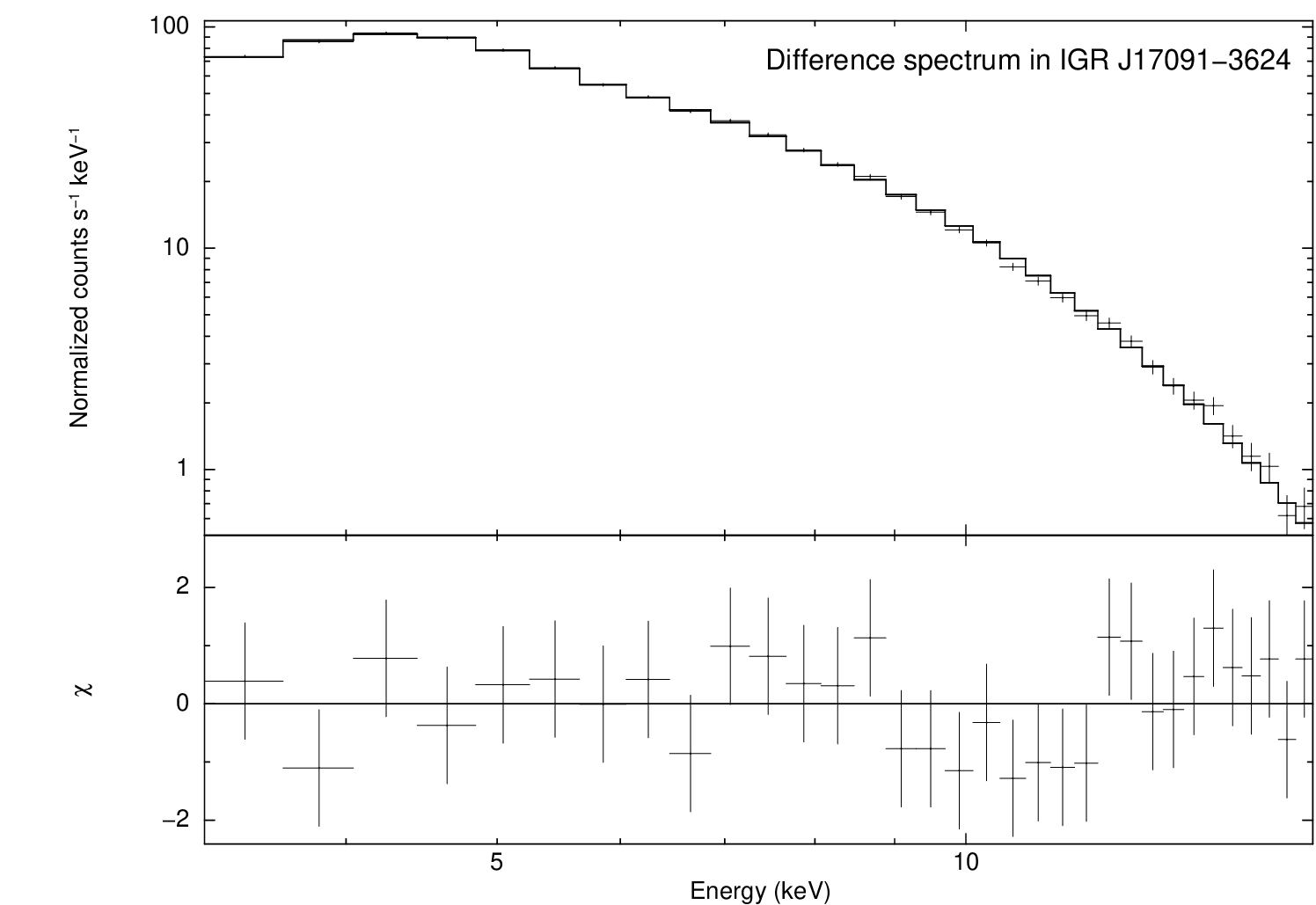} \\
\includegraphics[scale=0.30,angle=-90]{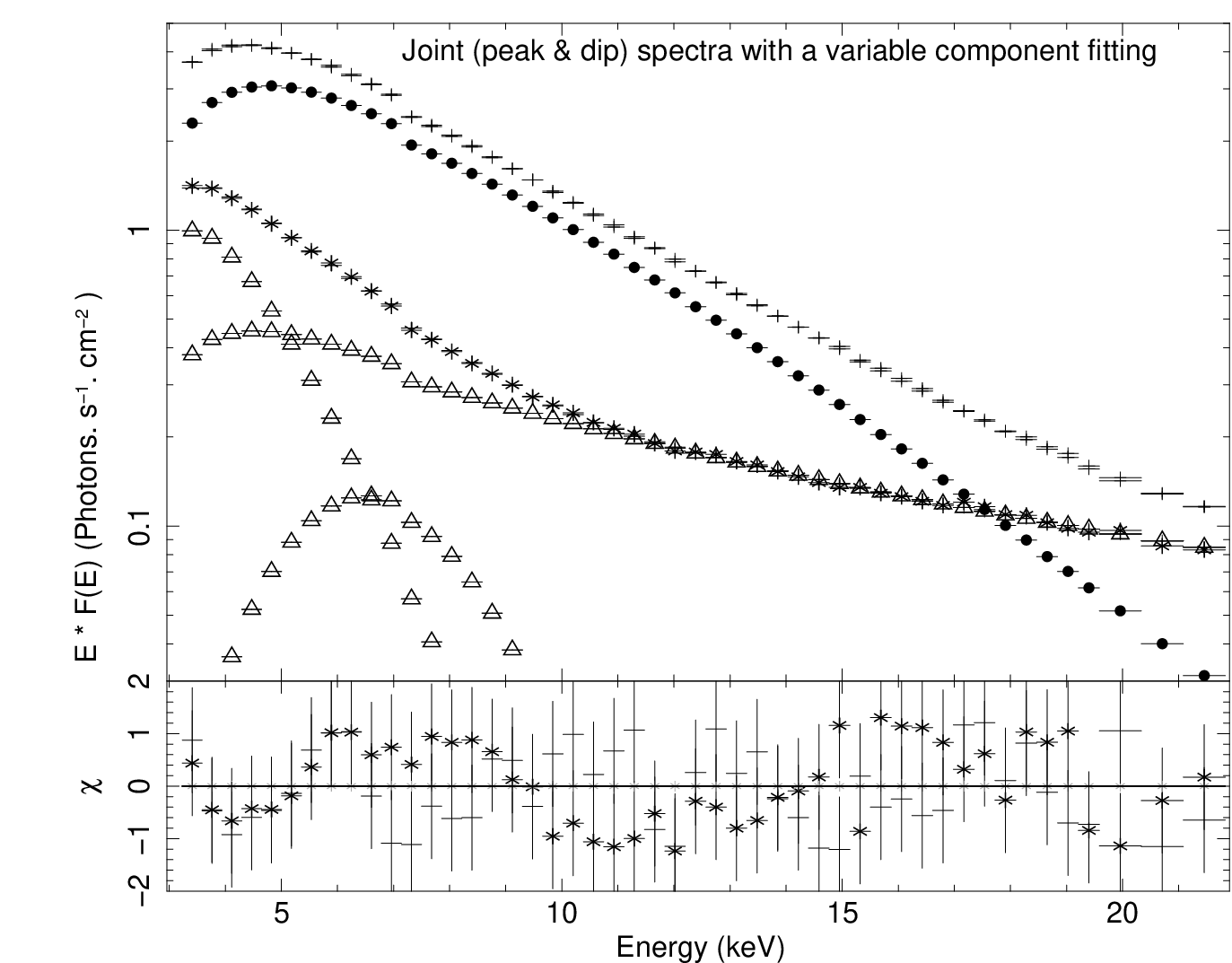} &
\includegraphics[scale=0.30,angle=-90]{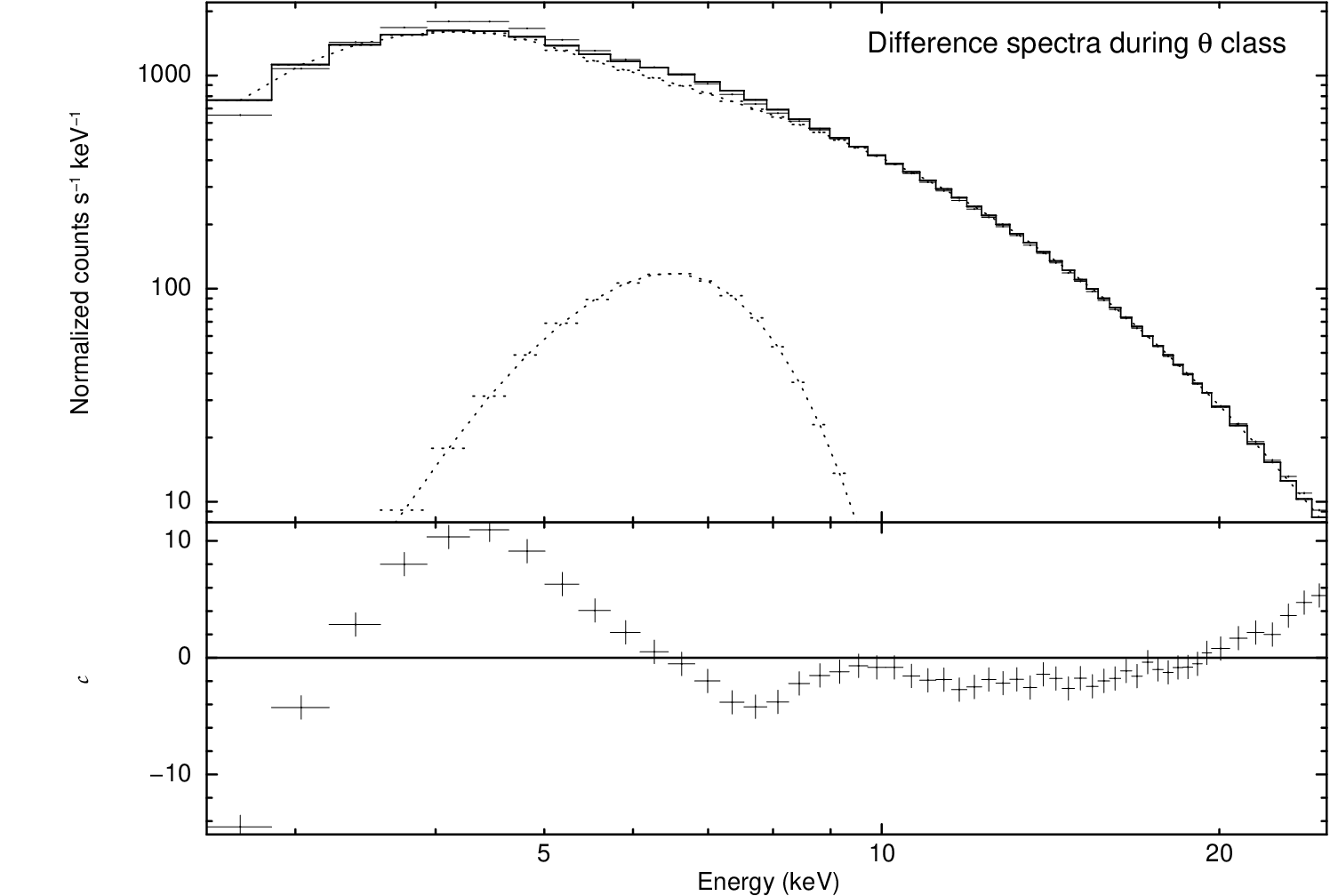} \\
 \end{tabular}
\caption{{\it Top$-$left and top$-$right panel:} Individual fitting of peak spectrum and dip spectrum with {\tt diskpbb+nthcomp+ga} model components along with their residuals in GRS 1915+105. {\it Middle$-$left and middle$-$right panel:} difference spectra fitted with the {\tt diskpbb+ga} model in GRS 1915+105 and {\tt diskpbb} model in IGR J17091$-$3624 respectively. {\it Bottom$-$left panel :} joint (peak and dip) spectra fitted with {\tt diskpbb+nthcomp+ga} model (tied) in GRS 1915+105 (model components are shown by triangles) plus a variable component ({\tt diskpbb}; shown in solid circles) are shown along with residuals. {\it Bottom$-$right panel:} an example of difference spectrum observed during $\theta$ class in GRS 1915+105, fitted with {\tt diskpbb+ga} model along with its residual. }
\end{center}
\end{figure*}

\section{Results}

Top$-$left panel of Figure 1 shows a typical {\it RXTE}/PCA light curve (1 s binning) of $\lambda$ class in the 2.0$-$60.0 keV energy range. Selected time intervals for extracting peak spectra and dip spectra are shown by red and green colour respectively. The position of peak flux level and persistent flux level are shown in the HID (middle$-$left panel of Figure 1) and in the CD (bottom$-$left panel of Figure 1). From the HID and CD, it may be noted that the highest count rate intervals for which peak spectra are extracted are generally described as the HS state/thermal dominated state \citep{b16,b56} and the lowest count rate intervals are described as the hard intermediate state (HIMS; \citet{b16,b56}) while intermediate states are shown in black colour \citep{b16,b56}. GRS 1915+105 never shows canonical low hard state as observed in other BHXBs \citep{b75}.

Similar to $\kappa$/$\lambda$ classes in GRS 1915+105, quasi-regular and large amplitude variations in X-ray intensity are also observed in IGR J17091$-$3624 by {\it RXTE}/PCA. The 2.0-60.0 keV light curve as observed on 19 April, 2011 is shown at the top right panel of Figure 1. From the light curve in IGR J17091$-$3624, it can be observed that the X-ray intensity is significantly lower and the length of peak and dip time intervals is significantly shorter than those of GRS 1915+105. Middle right and bottom right panels of Figure 1 show the HID and the CD of IGR J17091$-$3624 respectively as observed on 19 April, 2011. Although the HID and the CD of IGR J17091$-$3624 look different than that of GRS 1915+105, the red and the green colours in the HID and the CD in IGR J17091$-$3624 probably indicate the soft state and the hard state/HIMS respectively. 

\begin{figure*}
\begin{center}
 \begin{tabular}{c|c}
\includegraphics[scale=0.32,angle=-90]{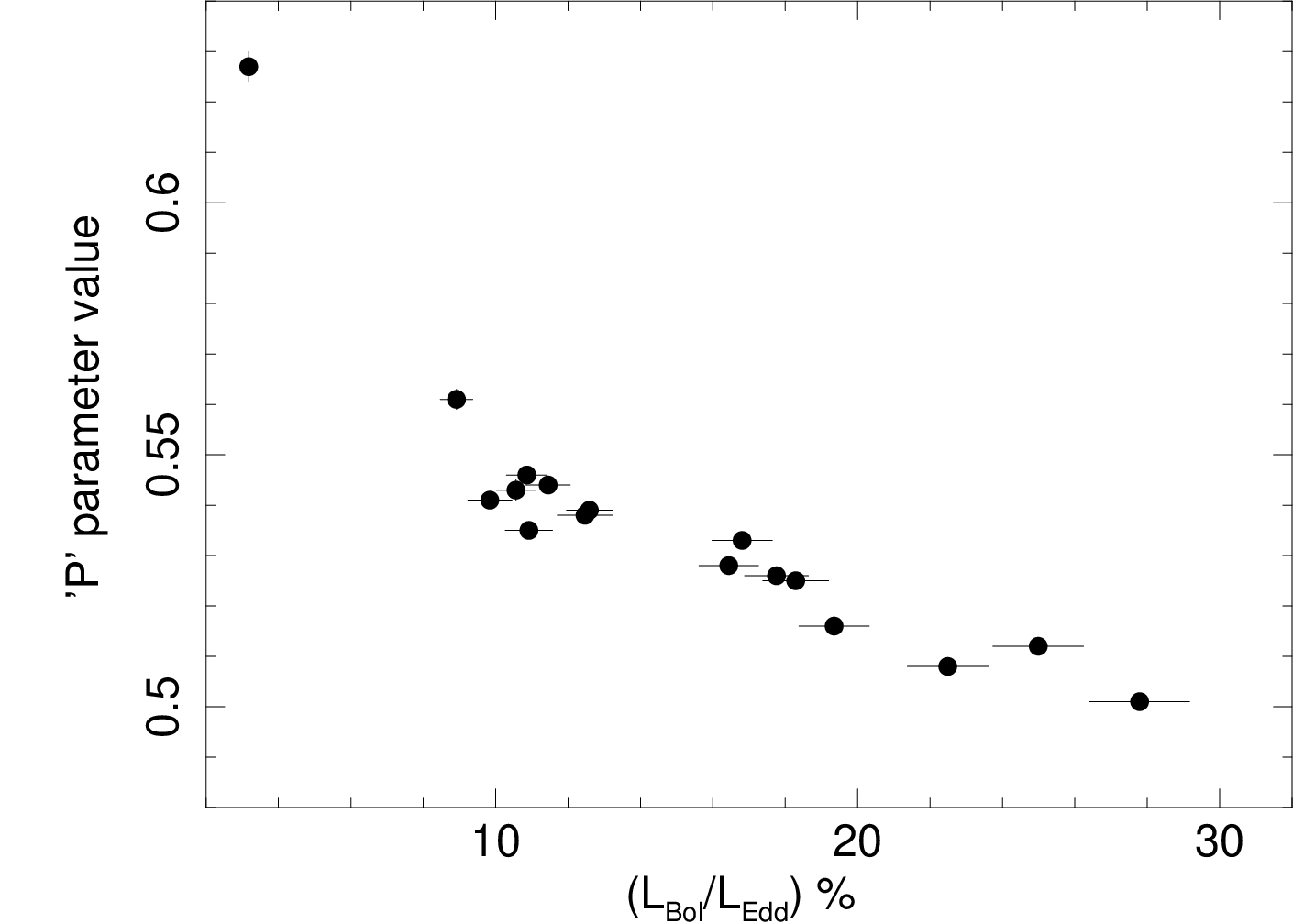} &
\includegraphics[scale=0.32,angle=-90]{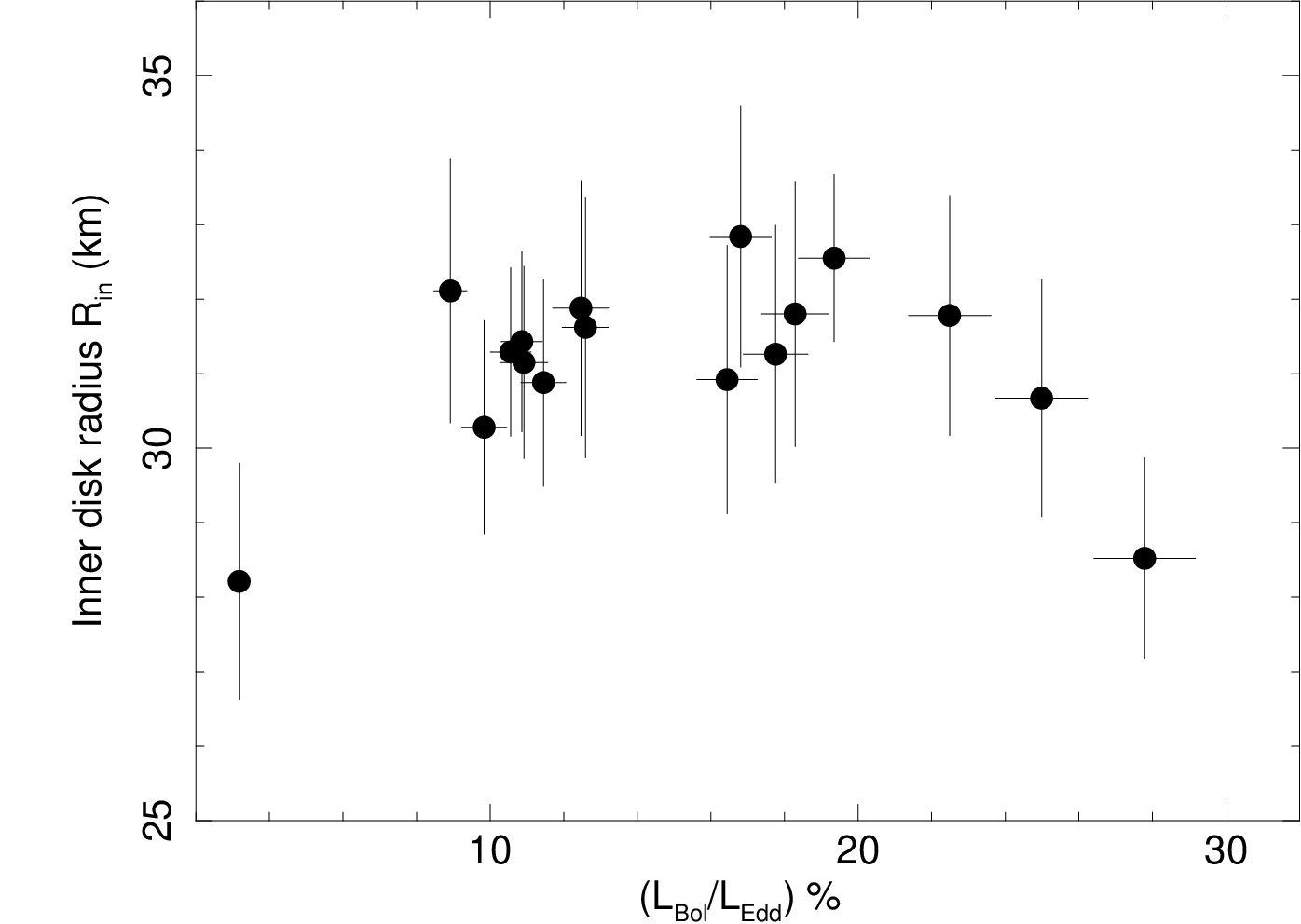} \\
\end{tabular}
\caption{{\it Left panel :} plot of `{\it p}' parameter value as a function of bolometric Eddington ratio, L$_{Bol}$/L$_{Edd}$ where L$_{Bol}$ is unabsorbed bolometric luminosity in the energy range 0.1$-$50.0 keV and L$_{Edd}$ is the Eddington luminosity for GRS 1915+105. {\it Right panel :} plot of inner disc radius R$_{in}$ (calculated using the distance of 12 $\pm$ 2 kpc and the inclination angle of 66$^0$) as a function of bolometric Eddington ratio, L$_{Bol}$/L$_{Edd}$ for GRS 1915+105. 1$\sigma$ error-bars are shown.}
\end{center}
\end{figure*}

 Top$-$left, top$-$right, bottom$-$left and bottom$-$right panels of Figure 2 show the squared rms variability in the energy range 2.0$-$4.7 keV, 4.8$-$6.4 keV, 6.5$-$8.6 keV and 8.7$-$10.9 keV respectively during both peak and dip intervals. These variabilities at all Fourier frequencies (0.01$-$10 Hz), show that the rms power during peak flux level is always higher than the rms power of persistent flux level. Even the shape of the variability during both intervals do not match with each other. This is an important observation from the variability analysis which support the idea that at all energy range, peak and dip spectra may be differed by the appearance and disappearance of an extra spectral component. Similar results and spectral parameters from both difference spectral parameter method and the joint spectra with a variable component (JSVC) method confirm that an additional component is indeed appeared during peak intervals. There are at least two more evidence that the additional component is more consistent with a {\it p}$-$free disc blackbody rather than a standard disc blackbody $-$ (1) the reduced $\chi^2$ averaged over all observations for ${\tt diskpbb}$ model is significantly better than the reduced $\chi^2$ averaged over all observations for ${\tt diskbb}$ model. The {\it F}$-$test yields the significance of 6.4$\sigma$. Again, from the JSVC method, it can also be noted that the choice of {\tt diskpbb} model over {\tt diskbb} model as a variable component is significant $>$ 5$\sigma$. (2) From the Table 2, it can be noted that for all observations, the `{\it p}' values during peak spectral fitting are significantly lower (at least 3$\sigma$) than the `{\it p}' values during dip spectral fitting. This indicate that the spectral modelling during peak intervals prefers slim disc approximations rather than standard disc prescriptions.     

Top$-$left and top$-$right panel of Figure 3 show the peak and dip spectra of GRS 1915+105, fitted with the {\tt diskpbb+nthcomp+ga} model respectively. Fitted model components and residuals are also shown. It may be noted that at energies less than 15 keV, the contribution of the {\tt diskpbb} model flux to the total flux decreases while moving from the peak spectral fitting (top left panel)  to the dip spectral fitting (top right panel). In all cases, the flux due to the Gaussian model is always $<$ 1\% of the total flux. Difference spectra from both GRS 1915+105 and IGR J17091$-$3624, fitted with the {\tt diskpbb} model are shown in the middle left and middle right panel of Figure 3 along with their residuals. Bottom left panel of Figure 3 shows joint spectral fitting of the peak (shown by $+$) and dip (shown by stars) spectra with the {\tt diskpbb+nthcomp} model (tied) plus the {\tt diskpbb} model as a variable component (shown by circles).

We may add here that during $\kappa$/$\lambda$ classes, a soft spectral component seems to appear during peak flux on the top of the hard persistent flux. However, during $\theta$ class, the behavior is opposite. The peak flux level is spectrally harder by a couple of factors than the persistent flux level, i.e., soft dips are observed. In this class, average count rate during short, soft dip intervals is $\sim$ 1749.32 cts$-$s$^{-1}$/PCU whereas average count rate during long, hard intervals is 4246.85 cts$-$s$^{-1}$/PCU (as observed on 05 September, 1997 (Obs ID $-$: 20402-01-45-02)).  Difference spectra are created by using the long, hard interval spectra as the source spectra in the {\tt XSpec v12.7.1} and the soft, dip interval spectra as its background spectra. Although \citet{b21} numerically reproduced the light curve of the $\theta$ class, we find that the difference spectra in the $\theta$ class is much more complex than the difference spectra in $\kappa$/$\lambda$ classes and cannot be described by a single component model like {\tt diskpbb} or {\tt nthcomp} along with the {\tt ga} at 6.4 keV. (reduced $\chi^2$ is greater than 5 for both models). Difference spectra due to the $\theta$ class fitted with {\tt diskpbb+ga} model is shown in bottom right panel of Figure 3. Hence the spectral behaviour during excess emission in the $\theta$ class is complex and is not reflected in the light curve. 

To check whether fitted parameters of difference spectra matches with simulated {\tt diskpbb} spectral parameters from earlier works \citep{b90}, we plot `{\it p}' parameter values (left panel of Figure 4) and apparent inner disc radius (right panel of Figure 4) as a function of the bolometric Eddington ratio. Both plots show the decrease in `{\it p}' parameter and apparent inner disc radius with the increase in bolometric Eddington ratio with the significance of 6.8$\sigma$ and 3.4$\sigma$, respectively. The observed trend in both plots agrees qualitatively (although not quantitatively) with the trend from numerical predictions from \citet{b90}.

\section{Discussion \& conclusion}
 
Using flux resolved spectroscopy, we show that during large amplitude, X-ray intensity variations in GRS 1915+105, the peak and the dip spectra of the source are complex and multi-components. After checking with results from several multi-component model fitting, we show that a combination of {\it p}$-$free disc model along-with a Comptonization model is more consistent ($>$ 6$\sigma$ significance using F-test) with the data than a standard disc model and Comptonization. More importantly, applying both difference spectral method (where the difference spectra are constructed by treating peak spectra as the source spectra in {\tt XSpec} and the dip spectra as its background spectra) and joint spectra with a variable component fitting method on several large amplitude oscillations from GRS 1915+105, we find that for all observations of $\kappa$ and $\lambda$ classes, the difference spectra is remarkably simple and is consistent with being a single {\it p}$-$ free disc emission. However, this may not hold true for complex variabilities where soft dips are observed (like $\theta$ class). 
We show that this is also the case for IGR J17091$-$3624. It may be noted that our analysis and interpretation of results are based on spectral modelling of the {\it RXTE}/PCA data. Although we put best efforts to bring out a physical model consistent with the data by eliminating possibilities of model degeneracies, it would be interesting to check the validity of our results using broad-band spectroscopy with {\it Swift}/XRT and {\it SUZAKU}. These active missions can provide spectra with larger energy range than {\it RXTE}.  

\begin{table*}
 \centering
 \caption{Spectral parameter values with 1$\sigma$ error-bars for GRS 1915+105 fitted with {\tt diskpbb} model using both difference spectral fitting and JSVC fitting methods. Epoch number of the corresponding observation is provided in the superscript with $\chi^2$/dof.}
\begin{center}
\scalebox{0.75}{%
\begin{tabular}{ccccccccccccccc}
\hline
& & & {\bf Difference}  & {\bf spectral} & {\bf fitting} & {\bf method}  & &  \vline & {\bf JSVC} & {\bf fitting} & {\bf method} \\
  Observation & Observation  & & & & & & & \vline & & & \\
\cline{3-8}
\cline{10-12}
 & date  & & & & & & & \vline & & & \\
no. & (MJD)  & kT$_{in}$ & R$_{in}$ & {\it `p'} parameter & Bolometric flux (L$_{bol}$) & L$_{bol}$/L$_{edd}$ & $\chi^2/dof$ & \vline & kT$_{in}$ & R$_{in}$ & {\it `p'} parameter \\
       &      & (keV) & (km) & value & (10$^{-8}$ ergs/s/cm$^{-2}$) &  (\%) &  & \vline & (keV) & (km) & value \\  
\hline
1 & 50363.27   & $2.39^{+0.05}_{-0.02}$ & $28.52^{+1.23}_{-1.44}$ & $0.501^{+0.001}_{-0.001}$ & $37.78^{+1.53}_{-1.95}$ & 27.46 & $45.56/43^{\bf 3}$ & \vline & $2.42^{+0.02}_{-0.03}$ & $30.45^{+4.23}_{-5.44}$ & $0.502^{+0.002}_{-0.001}$ \\
2 & 50363.34   & $2.33^{+0.05}_{-0.03}$ & $32.84^{+1.86}_{-1.65}$ & $0.533^{+0.001}_{-0.002}$ & $21.63^{+1.52}_{-1.97}$ & 16.18 & $48.56/43^{\bf 3}$ & \vline & $2.36^{+0.03}_{-0.04}$ & $35.21^{+2.21}_{-4.72}$ & $0.536^{+0.002}_{-0.003}$ \\
3 & 50363.41 & $2.32^{+0.05}_{-0.02}$ & $31.78^{+1.45}_{-1.78}$ & $0.508^{+0.001}_{-0.001}$ & $29.96^{+2.02}_{-1.81}$ & 22.05 & $46.16/43^{\bf 3}$ & \vline & $2.29^{+0.04}_{-0.03}$ & $34.65^{+1.92}_{-4.23}$ & $0.505^{+0.003}_{-0.002}$ \\
4 & 50627.55  & $2.41^{+0.06}_{-0.04}$ & $32.11^{+1.72}_{-1.85}$ & $0.561^{+0.003}_{-0.002}$ & $11.33^{+1.06}_{-1.34}$ & 8.43 & $41.29/43^{\bf 3}$ & \vline & $2.37^{+0.05}_{-0.03}$ & $38.45^{+2.25}_{-5.74}$ & $0.558^{+0.004}_{-0.003}$ \\
5 & 50627.62  & $2.44^{+0.05}_{-0.03}$ & $30.88^{+1.56}_{-1.24}$ & $0.544^{+0.003}_{-0.003}$ & $14.98^{+1.37}_{-1.26}$ & 11.02 & $47.31/43^{\bf 3}$ & \vline & $2.42^{+0.04}_{-0.04}$ & $32.11^{+3.31}_{-4.18}$ & $0.548^{+0.006}_{-0.005}$ \\
6 & 50639.63 & $2.47^{+0.03}_{-0.02}$ & $30.67^{+1.62}_{-1.55}$ & $0.512^{+0.002}_{-0.001}$ & $33.47^{+1.79}_{-1.57}$ & 24.61 & $49.57/43^{\bf 3}$ & \vline & $2.40^{+0.03}_{-0.04}$ & $35.23^{+2.13}_{-5.89}$ & $0.515^{+0.002}_{-0.004}$ \\
7 & 50639.74 & $2.46^{+0.06}_{-0.04}$ & $31.26^{+1.86}_{-1.61}$ & $0.526^{+0.002}_{-0.001}$ & $23.53^{+1.87}_{-1.42}$ & 17.26 & $44.01/43^{\bf 3}$ & \vline & $2.47^{+0.03}_{-0.03}$ & $30.36^{+2.58}_{-3.69}$ & $0.522^{+0.005}_{-0.003}$ \\
8 & 50641.49  & $2.46^{+0.04}_{-0.05}$ & $30.92^{+1.74}_{-1.88}$ & $0.528^{+0.002}_{-0.001}$ & $21.35^{+1.26}_{-1.27}$ & 16.01 & $45.98/43^{\bf 3}$ & \vline & $2.42^{+0.05}_{-0.03}$ & $33.49^{+2.96}_{-4.55}$ & $0.531^{+0.002}_{-0.004}$ \\
9 & 50641.56 & $2.44^{+0.04}_{-0.03}$ & $31.80^{+1.82}_{-1.75}$ & $0.525^{+0.002}_{-0.001}$ & $24.24^{+1.82}_{-1.73}$ & 17.91 & $40.19/43^{\bf 3}$ & \vline & $2.45^{+0.03}_{-0.03}$ & $35.56^{+2.15}_{-5.13}$ & $0.522^{+0.002}_{-0.003}$ \\
10 & 51284.16 & $2.37^{+0.04}_{-0.03}$ & $28.21^{+1.76}_{-1.41}$ & $0.627^{+0.003}_{-0.002}$ & $4.01^{+0.79}_{-0.57}$ & 3.17 & $43.67/38^{\bf 4}$ & \vline & $2.40^{+0.03}_{-0.03}$ & $31.25^{+1.86}_{-3.39}$ & $0.619^{+0.009}_{-0.007}$ \\
11 & 51284.09 & $2.42^{+0.04}_{-0.02}$ & $31.62^{+1.84}_{-1.62}$ & $0.539^{+0.002}_{-0.003}$ & $16.29^{+1.36}_{-1.15}$ & 12.36 & $40.18/38^{\bf 4}$ & \vline & $2.41^{+0.03}_{-0.02}$ & $34.87^{+3.41}_{-5.16}$ & $0.542^{+0.003}_{-0.004}$ \\
12 & 51284.28 & $2.41^{+0.05}_{-0.03}$ & $30.55^{+1.71}_{-1.68}$ & $0.516^{+0.002}_{-0.001}$ & $25.19^{+1.51}_{-1.33}$ & 19.06 & $41.84/38^{\bf 4}$ & \vline & $2.44^{+0.03}_{-0.04}$ & $33.28^{+2.25}_{-4.56}$ & $0.520^{+0.004}_{-0.004}$ \\
13 & 51284.35 & $2.40^{+0.05}_{-0.03}$ & $31.29^{+1.04}_{-1.13}$ & $0.543^{+0.003}_{-0.002}$ & $13.27^{+1.26}_{-1.37}$ & 10.18 & $39.27/38^{\bf 4}$ & \vline & $2.37^{+0.02}_{-0.03}$ & $35.69^{+1.89}_{-5.78}$ & $0.551^{+0.005}_{-0.006}$ \\
14 & 51284.41 & $2.41^{+0.05}_{-0.04}$ & $31.43^{+1.64}_{-1.39}$ & $0.546^{+0.003}_{-0.001}$ & $13.41^{+1.58}_{-1.23}$ & 10.34 & $42.01/38^{\bf 4}$ & \vline & $2.38^{+0.03}_{-0.03}$ & $33.29^{+2.47}_{-3.68}$ & $0.549^{+0.004}_{-0.003}$ \\
15 & 51284.49 & $2.39^{+0.06}_{-0.04}$ & $31.88^{+1.84}_{-1.56}$ & $0.538^{+0.003}_{-0.002}$ & $16.71^{+1.69}_{-1.63}$ & 12.03 & $38.35/38^{\bf 4}$ & \vline & $2.41^{+0.03}_{-0.04}$ & $30.53^{+3.64}_{-1.28}$ & $0.541^{+0.002}_{-0.002}$ \\
16 & 50617.61 & $2.46^{+0.03}_{-0.02}$ & $31.15^{+1.23}_{-1.36}$ & $0.535^{+0.002}_{-0.001}$ & $14.43^{+1.32}_{-1.68}$ & 10.51 & $46.28/43^{\bf 3}$ & \vline & $2.44^{+0.03}_{-0.04}$ & $34.45^{+2.26}_{-4.79}$ & $0.539^{+0.003}_{-0.004}$ \\
17 & 50617.54 & $2.50^{+0.04}_{-0.03}$ & $30.28^{+1.28}_{-1.66}$ & $0.541^{+0.002}_{-0.001}$ & $12.61^{+1.24}_{-1.58}$ & 9.58 & $47.12/43^{\bf 3}$ & \vline & $2.46^{+0.05}_{-0.03}$ & $32.98^{+2.46}_{-3.32}$ & $0.544^{+0.003}_{-0.003}$ \\
\hline
\end{tabular}}
\end{center}
\end{table*}

\begin{table*}
 \centering
 \caption{Spectral parameter values with 1$\sigma$ error-bars for IGR J17091$-$3624 fitted with {\tt diskpbb} model using both difference spectral fitting and joint spectra with variable component (JSVC) fitting methods. Epoch number of the corresponding observation is provided in the superscript with $\chi^2$/dof.}
\begin{center}
\scalebox{0.75}{%
\begin{tabular}{cccccccccccccc}
\hline
& & & & {\bf Difference}  & {\bf spectral} & {\bf fitting} & {\bf method}   &  \vline & {\bf JSVC} & {\bf fitting} & {\bf method} \\
  Observation & Observation & Observation  & & & & &  & \vline & & & \\
\cline{4-8}
\cline{10-12}
 & & date  & & & & & &  \vline & & & \\
no. & ID & ( MJD)  & kT$_{in}$ & Norm. & {\it `p'} parameter & Bolometric flux (L$_{bol}$) & $\chi^2/dof$ & \vline & kT$_{in}$ & Norm. & {\it `p'} parameter \\
    &   &      & (keV) &  & value & (10$^{-8}$ ergs/s/cm$^{-2}$) &   & \vline & (keV) &  & value \\  
\hline
1 &  96420-01-07-01 & 55662.73 & $2.03^{+0.06}_{-0.04}$ & $0.74^{+0.25}_{-0.47}$ & $0.501^{+0.001}_{-0.001}$ & $3.85^{+0.38}_{-0.76}$ & $27.49/28^{\bf 5}$ & \vline & $2.10^{+0.02}_{-0.04}$ & $0.92^{+0.29}_{-0.48}$ & $0.503^{+0.002}_{-0.003}$ \\
2 & 96420-01-07-00 & 55661.74 & $1.99^{+0.03}_{-0.02}$ & $0.75^{+0.33}_{-0.52}$ & $0.502^{+0.001}_{-0.001}$ & $3.49^{+0.22}_{-0.32}$ & $32.65/28^{\bf 5}$   & \vline & $2.04^{+0.02}_{-0.03}$ & $0.81^{+0.15}_{-0.39}$ & $0.505^{+0.003}_{-0.005}$ \\
3 & 96420-01-04-02 & 50363.41 & $1.87^{+0.12}_{-0.05}$ & $0.97^{+0.13}_{-0.24}$ & $0.502^{+0.001}_{-0.001}$ & $3.12^{+0.16}_{-0.14}$ & $31.14/28^{\bf 5}$  & \vline & $1.91^{+0.03}_{-0.05}$ & $0.85^{+0.14}_{-0.23}$ & $0.503^{+0.002}_{-0.003}$ \\
4 & 96420-01-05-04 & 50627.55 & $1.85^{+0.08}_{-0.04}$ & $0.84^{+0.21}_{-0.32}$ & $0.503^{+0.001}_{-0.002}$ & $2.96^{+0.19}_{-0.28}$ & $29.45/28^{\bf 5}$  & \vline & $1.82^{+0.04}_{-0.03}$ & $1.01^{+0.18}_{-0.44}$ & $0.504^{+0.001}_{-0.002}$ \\
\hline
\end{tabular}}
\end{center}
\end{table*}

In this work, our results answer a primary question regarding the nature of the large amplitude variations seen in these sources.
The variations could be associated with dramatic spectral changes, as seen when typical Galactic black holes
make transitions from hard to soft states or they could be due to quasi$-$periodic appearance and disappearance of
a spectral component. For the data analysed here, we find that the latter is a more consistent explanation. 

The absolute rms variability represents the intrinsic variability of any light curve. In order to have an extra$-$component, it is necessary (but not sufficient) to have excess power (may be small) at any energy and at any Fourier frequency of the absolute rms spectrum. If at any energy and at any Fourier frequency, the absolute rms of the PDS of the dip intervals was to exceed the absolute rms of the PDS of the peak interval, then the idea of an additional component would have to be abandoned. We verify that this is not so by noting that in four different energy bands, the short term rms variability of peak intervals is always higher than the dip intervals at different time-scales (0.01$-$10.0 Hz) and their shapes are significantly different from each other. It may be noted that the absolute rms could indicate a new component with more variability, but it could also indicate the brightening of a component with steady fractional variability. Thus, rms spectra analysis does not prove our interpretation but it is merely used for the consistency check only. 

Specifically the variable component can be described as a {\it p}$-$free disc emission which could arise if the
standard disc is super$-$Eddington, or if the colour factor varies with radii or if the disc is not steady. Using numerical simulations, \citet{b90} show that the deviation from the standard accretion disc (i.e. T$_{in}$ $\propto$ R$_{in}^{-3/4}$) to the optically thick slim disc (i.e. T$_{in}$ $\propto$ R$_{in}^{-1/2}$) becomes significant when the source accretes matter at Eddington or super$-$Eddington limit (see bottom panel of Figure 5 in \citep{b90}). Comparing Table 5 and Figure 4 from our analysis with Figure 5 in \citet{b90}, we find that the observed values of T$_{in}$ and R$_{in}$ from our spectral analysis of GRS 1915+105 agrees well with simulated values from \citet{b90}. {\it `p'} parameter values as well as R$_{in}$ show correct trend (Figure 4) as observed by \citet{b90}. In all cases, as luminosity increases, both {\it `p'} parameter values as well as R$_{in}$ decreases.
For the difference spectra the index p varies inversely with luminosity as expected if the disc is
super$-$Eddington. However, the bolometric luminosity itself is just $\sim 20 $-$ 30$\% of Eddington luminosity, which suggests that
the other reasons quoted may also be the reason for the deviation from the standard disc. Indeed, Using both high ($\delta$ class) and low luminosity ($\chi_C$ class) observations from GRS 1915+105, \citet{b80} showed that for the spin parameter value of 0.99, the disc luminosity prefers a flatter temperature gradient than that in the standard disc when the luminosity is above $\sim$ 0.2L$_{EDD}$ (see Figure 11). Later, in an extensive study, \citet{b82} derive equations governing the time evolution of fully relativistic slim accretion discs in the Kerr metric and numerically construct their detailed non$-$stationary models. Using these models, they also predicted that the disc luminosity of a moderately rotating black hole, deviates from the standard disc assumption at $\sim$0.2L$_{EDD}$ or above. We are considering here the bolometric luminosity due to difference spectrum which is basically the change in luminosity when the source evolves from the persistent flux level to the peak flux level. The peak level luminosity may reach 0.6$-$0.8 L$_{EDD}$. If this hypothesis is correct, then the slim disc may be able to describe the accretion disc during large amplitude X-ray variations observed during $\kappa$ or $\lambda$ class in GRS 1915+105 and IGR J17091$-$3624. 

Our results are consistent with the general idea that the variability is due to an unstable radiation pressure
dominated disc. Such discs should show high amplitude oscillations where the difference between the peak and dip flux
may be more than an order of magnitude. If there is an additional persistent source in the system, the oscillations
will be observed as an appearance and disappearance of a disc like emission over the persistent level and we show that this is
precisely what is observed. Although \citet{b21} numerically reproduce oscillations with equal time in peaks and dips using disc-corona coupled system, it would be interesting to see whether detailed simulations of the radiation pressure dominated discs could reproduce such oscillations.

The model requires a persistent flux which itself maybe described as some soft thermal disc emission and a hard 
Comptonized emission. The origin of this is rather unclear. It is unlikely that its origin could be the outer disc (R $>$ 300R$_g$) as the dynamical structure associated with the outer disc cannot be changed on the limit-cycle time-scale which is much shorter than the viscous time$-$scale of the outer disc. Again, the luminosity of persistent intervals is too high ($\sim$ 5\% of L$_{Eddington}$) to come from the  outer disc. \citet{b95} showed that evolution of the inner disc geometry which is in accordance with the theoretical interpretations of slim disc approximations \citep{b89,b11} and numerical simulations \citep{b12}, may be related to the dramatic drop in luminosity during large amplitude oscillations. This interpretation is supported by our results from the spectral analysis using both methods. Detail simulations of the radiation pressure dominated disc are required to see whether the outer gas pressure dominated disc as well as jet activity can be reconciled with the origin of the persistent emission.  

Spectral results show that the {\tt diskpbb} model provides low inner disc radius and high disc temperature. This is highly consistent with the black hole properties of GRS 1915+105. As pointed out by \citet{b90}, there are two possible explanations of apparently low inner disc radius and high disc temperature in GRS 1915+105 $-$ (1) the presence of a highly rotating Kerr black hole in GRS 1915+105 \citep{b79} with the spin parameter of $\sim$ 0.7-0.98 \citep{b76,b78,b77}. The dragging of last stable orbit inward is an obvious consequence of such high spin which in turn increases the inner disc temperature and decreases the inner disc radius. This fact agrees well with our values of R$_{in}$ and inner disc temperature. (2) Due to large inclination of GRS 1915+105 ($\sim$70$^o$), the outer part of the optically thick disc can effectively block the radiation from the inner part which effectively reduce the value of R$_{in}$. 

The other BHXB IGR J17091$-$3624 shows lower average disc temperature than GRS 1915+105 and marginally low {\it `p'} parameter value. Although the unabsorbed bolometric flux is $\sim$ 20$-$50 times lesser than that of GRS 1915+105, our results show that the large amplitude X-ray variations observed in IGR J17091$-$3624 are also consistent with the slim disc approximations. Comparing results from both sources from this work and earlier works \citep{b83,b54,b53}, it may be possible that the nature of the binary companion in both sources is same and X-ray activities in these sources are related by a scaling of the central black hole mass. However, this indeed need to be investigated further. Since mass, distance (but a tentative measurement of large distance has been reported in IGR J17091$-$3624 by \citet{b53}) and spin parameters are unknown in IGR J17091$-$3624, further prediction of black hole characteristics in IGR J17091$-$3624 is presently out of scope. 

\section*{Acknowledgements}

We would like to thank the anonymous referee for his constructive comments and suggestions which help to improve the paper. This research has made use of data obtained through the High Energy Astrophysics Science Archive Research Center online service, provided by the NASA/Goddard Space Flight Center.

\bsp

\label{lastpage}

\end{document}